\documentclass{article}

\usepackage{amsmath,amssymb}
\usepackage{PRIMEarxiv}
\usepackage[utf8]{inputenc} 
\usepackage[T1]{fontenc}    
\usepackage{hyperref}       
\usepackage{url}            
\usepackage{booktabs}       
\usepackage{amsfonts}       
\usepackage{nicefrac}       
\usepackage{microtype}      
\usepackage{lipsum}
\usepackage{graphicx}       
\graphicspath{{media/}}     
\usepackage{cleveref,textcomp}
\usepackage{cite,enumitem}
\usepackage{algorithmic}
\usepackage[table,xcdraw]{xcolor}
\usepackage{xcolor}
\usepackage{subfig}
\newtheorem{theorem}{Theorem}

\usepackage{color}
\usepackage{mathrsfs}

\newtheorem{definition}[theorem]{Definition}

{ 
\newtheorem{remark}[theorem]{Remark}

}

\usepackage{wrapfig}
\usepackage{float}

\newcommand{\bi}{\begin{itemize}}
\newcommand{\ei}{\end{itemize}}
\newcommand{\bd}{\begin{displaymath}}
\newcommand{\ed}{\end{displaymath}}
\newcommand{\be}{\begin{eqnarray*}}
\newcommand{\ee}{\end{eqnarray*}}

\newcommand{\eg}{\textit{e}.\textit{g}.}
\newcommand{\ie}{\textit{i}.\textit{e}.}
\newcommand{\eetitle}[1]{\noindent{\underline{\em #1}}}
\newcommand{\etitle}[1]{\noindent{\underline{ #1}}}
\newcommand{\stitle}[1]{\noindent{\bf #1}}

\title{Graph Neural Network and Koopman Models for Learning Networked Dynamics: A Comparative Study on Power Grid Transients Prediction
\thanks{The Pacific Northwest National Laboratory (PNNL) is operated by the Battelle Memorial Institute for the U.S. Department of Energy under Contract No. DE-AC05-76RL01830\,. The authors gratefully acknowledge the funding support from the Data Model Convergence (DMC) Initiative at PNNL. 
} 
}

\author{
  Sai Pushpak Nandanoori$^{1}$, Sheng Guan$^{2}$, Soumya Kundu$^{1}$, Seemita Pal$^{1}$, Khushbu Agarwal$^{1}$, \\ \textbf{Yinghui Wu}$^{2}$, \textbf{Sutanay Choudhury}$^{1}$ \\
  $^{1}${Pacific Northwest National Laboratory} \\ Richland, WA 99354, USA \\ 
  (emails: \texttt{\{saipushpak.n, soumya.kundu, seemita.pal,} \\ 
  \texttt{khushbu.agarwal, sutanay.choudhury\}@pnnl.gov)} \\
  $^{2}${Case Western Reserve University} \\
  Cleveland, OH 44106, USA \\ 
  (emails: \texttt{\{sxg967, yxw1650\}@case.edu)}
}

\begin{document}
\maketitle

\begin{abstract}
Continuous monitoring of the spatio-temporal dynamic behavior of critical infrastructure networks, such as the power systems, is a challenging but important task. In particular, accurate and timely prediction of the (electro-mechanical) transient dynamic trajectories of the power grid is necessary for early detection of any instability and prevention of catastrophic failures. Existing approaches for prediction of dynamic trajectories either rely on the availability of accurate physical models of the system, use computationally expensive time-domain simulations, or are applicable only at local prediction problems (e.g., a single generator). In this paper, we report the application of two broad classes of data-driven learning models -- along with their algorithmic implementation and performance evaluation -- in predicting transient trajectories in power networks using only streaming measurements and the network topology as input. One class of models is based on the \textit{Koopman operator theory} which allows for capturing the nonlinear dynamic behavior via an infinite-dimensional linear operator. The other class of models is based on the \textit{graph convolutional neural networks} which are adept at capturing the inherent spatio-temporal correlations within the power network. Transient dynamic datasets for training and testing the models are synthesized by simulating a wide variety of load change events in  the IEEE 68-bus system, categorized by the load change magnitudes, as well as by the degree of connectivity and the distance to nearest generator nodes. The results confirm that the proposed predictive models can successfully predict the post-disturbance transient evolution of the system with a high level of accuracy.
\end{abstract}

\stitle{\textsc{Notice of Publication:}} This work is currently under review in a journal. 

\section{Introduction}
Critical infrastructure networks, such as traffic networks and power systems, are often modeled as networked dynamical systems. Accurate learning of the spatio-temporal dependencies of the node-level attributes, which constitute the state of the dynamical network, is essential to monitoring and control of such systems. Power systems, for example, are increasingly operated closer to their stability limits (e.g., high loading) \cite{sobbouhi2021transient}, which, when coupled with the fast timescales associated with the electro-mechanical transients, makes timely and accurate prediction of trajectories absolutely critical to energy security and resilience. Enabling accurate predictions of the transient dynamic states  help detect the possible state violations into the future and thereby help predict possible line or link failures which when allowed to persist may lead to cascading failures. Moreover, emerging data-driven control techniques in power systems, such as deep reinforcement learning \cite{glavic2017reinforcement,huang2019adaptive}, rely on accurate time-series prediction of trajectories for the learning and exploration tasks. Even though modern power network comprises of various sensors, such as the phasor measurement units (PMUs), continually relaying measurements to the control center, robust prediction of transient dynamic trajectories remains challenging due to its enormity, complexity and nonlinearity.

\stitle{State-of-the-art}. Data-driven learning of predictive models for nonlinear dynamical systems has received some attention in recent times \cite{legaard2021constructing,wang2017new,ogunmolu2016nonlinear}. The work \cite{legaard2021constructing} presents a review of the different neural network-based learning models for nonlinear dynamical systems, including different variants of recurrent networks, e.g., long-short-short-term memory (LSTM) models \cite{wang2017new,ogunmolu2016nonlinear}. 
In the context of power systems dynamics, transient stability assessment and prediction has been an active area of research over the years, as summarized in a recent review \cite{sobbouhi2021transient}. It was noted that time-domain simulation-based approaches often yield the most accurate results, but are burdened by extreme computational costs. On the other hand, machine learning (ML) methods were found to be used most frequently, even though their applications were largely limited to the transient stability classification problem (yes/no question), as opposed to time-series prediction. In an early ML application for time-series prediction of transients, \cite{bahbah2004new} uses recurrent radial basis functions and artificial neural networks for generators’ angles and angular velocities prediction, albeit in small networks. LSTM has been used in \cite{li2020machine} for time-series prediction of post-contingency trajectories, but only for a single generator. {In} \cite{liu2019transient}, {a transient stability prediction method based on deep belief network and LSTM network is developed to predict generator rotor angle trajectories in the IEEE 39-bus network. However, this predictive model relies on an extensive list of input features (including detailed generator model parameters) which are often unavailable, unreliable, or expensive to identify} \cite{wu2020model}. {In contrast, the methods proposed in this work rely only on streaming PMU measurements (that includes frequency and voltage magnitudes).} 

We highlight here the difference between the task of multi-timestep prediction of transient trajectories, as covered in this work, and the problem of dynamic state estimation which has received some attention in the power systems community \cite{zhao2020roles}. In dynamic state estimation, the states of the system at the current time-point are estimated using available measurements and some knowledge of the system model. In contrast, this article is concerned with the time-series prediction of the transient trajectories over a look-ahead period (of a few seconds) using available measurements, without any prior knowledge of the system model.

\stitle{Enabling Models}. As explained above, the existing ML methods for transient prediction either place strong assumptions on the availability of extensive system features (\eg, detailed generator models), or limit themselves to only local prediction problems (\eg  transient trajectories of a single generator), and hence are not applicable to system-wide transient prediction problems under limited information (\eg access to noisy measurements data only). In this article, we focus on data-driven learning models that rely only on streaming measurements from the sensors (e.g., PMUs) across the network and, perhaps, some limited information such as the topology of the network and the locations of the sensors. Specifically, we choose two classes of learning models -- one based on the Koopman operator theory and the other based on the graph convolutional neural networks -- which we demonstrate to have the expressive capability to encode the nonlinearity in power system transient dynamics, and capture the dynamic behavior across the network, without requiring detailed physical models of the system. The following are the two major enabling techniques considered in this article for robust transient prediction:
\begin{enumerate}[leftmargin=0.4cm,label=\textbullet]
    \item predictive models that incorporate {\em Koopman operators} to learn the complex nonlinear dynamic behavior of power system transients in the form of infinite-dimensional linear maps in some \textit{lifted} space of observables; and 
\item {\em graph convolutional neural networks} that model power systems as dynamic graphs, and exploit their spatial-temporal-topological correlation for accurate prediction. 
\end{enumerate}

The choice of these two methods allows us to perform a comparison between predictive models that attempt to learn the underlying nonlinear dynamics governing the evolution of system trajectories, and those that learn the inherent spatio-temporal correlations in a dynamical network. In what follows, we provide brief summaries of the applications of the above two predictive models in related problems.

\eetitle{Koopman Operator Theory}.  
The Koopman operator theory (KOT) introduced in \cite{koopman1931hamiltonian}, represents the nonlinear dynamical system as a globally linear system on a space of (observable) functions. 
The evolution of these observables is given by the Koopman operator, a linear, infinite dimensional operator. Finding such an infinite-dimensional operator for generic nonlinear systems is computationally intractable. Nevertheless, several efficient approaches have been developed to construct a {\em finite-dimensional approximation} of the Koopman operator, including {\em dynamic mode decomposition (DMD)} \cite{rowley2009spectral,tu2013dynamic}, {\em extended DMD} \cite{williams2015data}, {\em robust DMD} \cite{sinha2018robust}, and {\em deep DMD (deepDMD)} \cite{takeishi2017learning,li2017extended,lusch2018deep,yeung2019learning}. 
Koopman operator-based techniques have found many applications in analysis of nonlinear systems, especially in fluid mechanics \cite{rowley2009spectral}, synthetic biology \cite{yeung2019learning,nandanoori2020data,sinha2020koopman} and energy systems \cite{eisenhower2010decomposing,barocio2014dynamic,susuki2016applied,nandanoori2020model,surana2018koopman,yeung2019learning,sinha2020data}. Majority of the data-driven KOT applications in power systems have focused on nonlinear modal analysis (via the use of Koopman modes), \eg coherency detection \cite{barocio2014dynamic,susuki2016applied}, attack identification \cite{nandanoori2020model} or anomaly diagnosis \cite{surana2018koopman}, with a few also considering time-series prediction  \cite{surana2018koopman,yeung2019learning,netto2021analytical}. The work in \cite{surana2018koopman} used DMD for transient prediction in power systems, assuming simpler swing-dynamic models of the generators and access to full-state measurements, both of which do not generalize to real-world systems. The work in \cite{yeung2019learning}, on the other hand, combined KOT with deep learning tools to develop a deepDMD model for power systems transient prediction and demonstrated its advantages over EDMD, albeit on a test system using simpler swing-dynamic models of the generators and full-state measurements. In a more recent work on analytical construction of KOT-based models \cite{netto2021analytical}, authors only consider a single-machine infinite-bus example and do not address the feasibility of the approach (\eg in terms of accuracy and scalability) to realistic multi-machine networks. Related works have also explored the use of KOT-based predictive models for feedback control design \cite{korda2018power}.

\eetitle{Graph Neural Networks}.
An alternative option to tackle the non-linearity and spatial-temporal dynamics of power systems is to exploit the emerging graph neural networks (GNNs) \cite{scarselli2008graph,zhou2020graph}. {Generally speaking, the GNNs are a class of deep models that represent a graphical network as node embeddings and learn the best representation 
(by optimizing learning objective function) that can be used for downstream tasks. 
In contrast to the conventional 
predictive models, such as recurrent neural networks (RNNs) and convolutional neural networks (CNNs)}\cite{lecun2015deep},
{the GNNs are able to jointly characterize node and topological features and thus have been adopted for network learning. CNNs are known to be limited to the regular Euclidean data computation, where the localized convolutional filters and pooling operators are difficult to be defined in the non-Euclidean power-grid data. RNNs on the other hand, are not optimized to process graphs, as they require to stack feature of nodes in a specific order, whereas there exists no natural order in  power grid data. In} \cite{yu2017spatio}, {a novel GNN model named the Spatio-Temporal Graph Convolutional Network (STGCN) was developed for mid/long-term time-series forecasting of traffic networks. Specifically, instead of applying regular convolutional and recurrent units, the STGCN deployed a fully convolutional structure on the time axis to extract spatio-temporal features from graph-structured traffic data, resulting in faster training with fewer parameters.} Application of GNN to power systems thus far, however, has been sparse. In a recent work \cite{lin2021spatial}, though, the authors used a GNN model for short-term (hours to days) load forecasting in residential buildings, and demonstrated its capability to capture the latent spatial dependencies of different houses within some geographic area.
 
\stitle{Contributions}. 
The main contributions of the paper are that:

\begin{enumerate}[leftmargin=0.2in,label=\arabic*)]

\item We construct data-driven variants of KOT-based models (namely, \textit{deepDMD} and \textit{robust DMD}) for multi-step prediction of the power system transients, using only a single time-step measurement. The KOT-based predictive models learn the inherent nonlinear dynamics purely from available measurements, without any prior information of the physics-based system models. {To the best of authors}\textquotesingle \; {knowledge, this is the first time that the transient predictions on a large power system (IEEE 68 bus) has been studied where the synthetic data is obtained using higher-order synchronous generator models (eighth order) and KOT-based methods have been applied using only critical measurements such as frequency and voltage. Only the work in } \cite{yeung2019learning} {among the existing works discusses the transient predictions however, on a small power network (IEEE 39 bus) with simplified swing equation model for the synchronous generators (second order).  }

\item In a novel application, we implement, train, and test a deep STGCN model (coupled with an \textit{incremental inference} strategy) for multi-step transient prediction in power systems. With the knowledge of the sensor measurements and the connectivity between the sensor nodes, the STGCN jointly learns the underlying temporal and spatial features for predicting the transient trajectories. {It is important to mention that the STGCN model (in the literature} \cite{yu2017spatio}) {is originally introduced for one-time step traffic prediction only. }

\item {The synthetic spatio-temporal time-series measurements which corresponds to different load disturbances of the IEEE 68 bus system are generated as part of this work using GridSTAGE } \cite{gridstage2020} {and are made accessible to the wider community} \cite{nandanoori2021nominal}. {Moreover, the KOT-based methods and STGCN model development codes developed as part of this research are also made accessible online} \cite{gridprediction2021} {for reproducible research.} 

\item Detailed comparative evaluation of the data-driven learning models are performed to demonstrate the feasibility of their application to power network transient prediction. Specifically, numerical studies are carried out to evaluate the robustness to noisy measurements, accuracy across spatio-temporal scales, sensitivity to hyper-parameters, and computational time costs. 
\end{enumerate}
 
The rest of the paper is structured as follows. Sec.\,\ref{sec-kot} and Sec.\,\ref{sec-gnn} provide brief technical description of the KOT-based and the GNN-based models, respectively. Sec.\,\ref{sec:setup} describes the simulation environment, the procedures for generating synthetic datasets, and the performance metrics. The numerical results from the detailed performance evaluation of the predictive models are presented in Sec.\,\ref{sec:results}. Finally, we conclude the article in Sec.\,\ref{sec-concl}, along with a discussion of the future directions.

\vspace{-1ex}
\section{Koopman Operator-Based Prediction}
\label{sec-kot}

We start with a revisit of the mathematical background for KOT (see  \cite{mezic2013analysis,williams2015data,lasota2013chaos} and the references therein).  
Consider the following discrete-time nonlinear system
\begin{align}
Z_{t+1} = F(Z_t),
\label{eq:discrete_dyn_sys}
\end{align}
where $Z_t\in \mathcal{M}\subseteq\mathbb{R}^m$ are the $m$-dimensional system states, and $F: \mathcal{M} \to \mathcal{M}$ is a continuously differentiable system map. Note that, in this work, we consider the map $F(\cdot)$ to be unknown.
We denote by $\mathcal{F}(\mathcal{M})$ the space of all scalar-valued functions of the system states $\pmb z$. 

\begin{definition}(Koopman Operator)
The Koopman operator, $\mathbb{U}: \mathcal{F}(\mathcal{M}) \to \mathcal{F}(\mathcal{M})$, associated with the dynamical system \eqref{eq:discrete_dyn_sys} is defined as an infinite-dimensional linear operator which acts on any scalar-valued observable function $\phi\in \mathcal{F}(\mathcal{M})$ of the system \eqref{eq:discrete_dyn_sys} as follows:
\begin{align}\label{eq:koopman}
[\mathbb{U} \phi](\pmb z) = \phi(F(\pmb z)).    
\end{align} 

\end{definition}
The evolution of observables is given by the Koopman operator, $\mathbb{U}$,  which is linear and the function space, $\mathcal{F}(\mathcal{M})$ is invariant under the action of the Koopman operator. Moreover as the space of functions is infinite dimensional, the Koopman operator which maps these functions is also infinite-dimensional, and this process of mapping is usually referred to as \textit{lifting}. Koopman operator is also a positive operator such that for any $\phi(\pmb z) \geq 0$, we obtain $[\mathbb{U}\phi] (\pmb z) \geq 0$ \cite{lasota2013chaos}. 

\stitle{Remark}. In the case where the mathematical model \eqref{eq:discrete_dyn_sys} is not known explicitly, a Koopman operator can be identified from sampled time-series data obtained from experiments or sensor measurements. 

In practice, a finite dimensional approximate of the Koopman operator is computed. In the scope of this work, Robust DMD (that results in the dimension of the approximate Koopman operator equal to the number of measurements) and deepDMD (that results in the Koopman operator whose size depends on the chosen observable space) are implemented to build predictive models for the power network from PMU data. 

\stitle{DeepDMD and Robust DMD}.
We adapt the deepDMD model introduced in \cite{yeung2019learning} and extend it to the case where partial measurements are available at each bus. While \cite{yeung2019learning} assumes full-state measurements (i.e., all the dynamic states at each bus of the system are measured), this present work assumes a more realistic scenario in which only a subset of states at any bus are measured. Let us denote by $X_t\in\mathbb{R}^n$ the $n$-dimensional measurements (e.g., frequencies and voltages) available at any time-instant $t$ from the PMUs installed in the power network. Henceforth, we will assume to have access to only the measurements $X_t$, and not the underlying system states (nor the system model). 

We define $\pmb{\psi}:\mathbb{R}^n \to \mathbb{R}^q$ as a $q$-dimensional vector-valued observable function of the measurements $X_t$ as follows:
\begin{align*}
\pmb{\psi}(X_t):=\begin{bmatrix}\psi_1(X_t) & \psi_2(X_t) & \cdots & \psi_q(X_t)\end{bmatrix}^\top
\end{align*}
where each $\psi_i$ ($i=1,2,\dots,q$) is a scalar-valued observable function of the measurements $X_t$\,. 

\etitle{Learning Objective}. 
The objective of the deepDMD is to learn such a mapping $\pmb{\psi}$ and the associated linear operator (the approximate Koopman) in \eqref{eq:koopman} which governs the evolution of the observable functions on the \textit{lifted} observables-space. While the observable function $\pmb \psi$ maps the time-series data to the observable space, we also need an \textit{inverse mapping} to retrieve the measurements from the observable space and this can be achieved in two different ways. The first one includes identifying the function $\pmb \varphi\equiv\pmb\psi^{-1}$ such that we have $X_t = \pmb \varphi (\pmb\psi(X_t))$\,. The second one involves appending the actual measured states ($X_t$) with observables to obtain \textit{measurement-inclusive observables}, defined as below:
\begin{align*}
    \pmb \Psi(X_t) = \begin{bmatrix} X_t \\ \pmb \psi(X_t)
    \end{bmatrix}.
\end{align*}
Since the measurements are accessible directly, there is no need for an inverse mapping in the case of measurement-inclusive observable functions. Due to its simplicity, we adopt this second approach of constructing measurement-inclusive observables in this paper. Note its similarity to the notion of \textit{state-inclusive} observables in \cite{yeung2019learning,johnson2018class}.

\begin{figure*}[h]
    \centering
    \includegraphics[scale = 1.1]{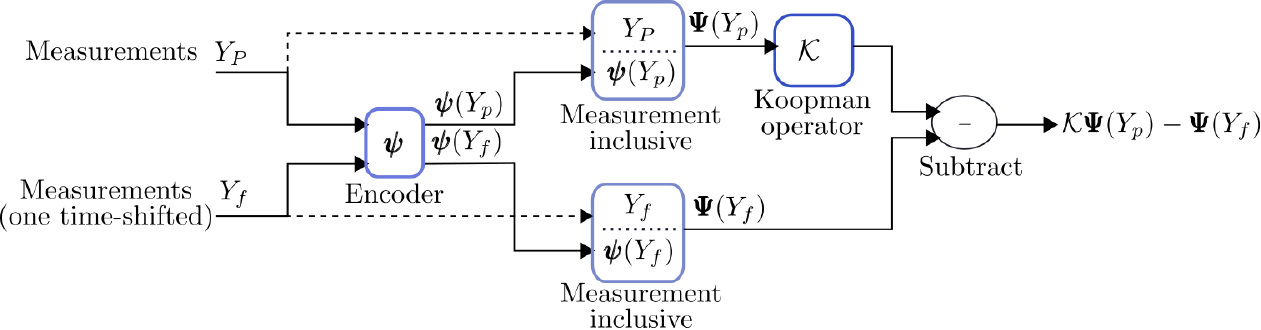}
    \caption{Computational overview for deepDMD.}
    \label{fig:deepDMD_overview}
\end{figure*}

The learning task involves identifying a suitable observable function $\pmb{\Psi}$, and a corresponding Koopman operator $\mathcal{K}$ such that $\pmb\Psi(X_{t+1})=\mathcal{K}\pmb\Psi(X_t)$, at all time-points $t$\,. Since in deepDMD, this learning task is performed via a neural network, we denote the observable as $\pmb{\Psi}(X_t,\Theta)$ (instead of $\pmb{\Psi}(X_t)$), where $\Theta$ represents the neural network weights and biases. 
Consider any sequence of $N+1$ measurement snapshots $\left[{X}_0, X_1, \dots, X_{N-1},X_N \right]$. Let us define the following:
\begin{align*}
    Y_p &:= \left[X_0, \,X_1, \,\dots, \,X_{N-2},\,X_{N-1} \right], \\
    Y_f &:= \left[X_1, \,X_2,\, \dots, \,X_{N-1},\,X_N \right]. 
\end{align*}
where both $Y_p$ and $Y_f$ are a collection of measurements from $N$ time-points, and $Y_f$ is constructed from $Y_p$ by shifting the measurements by one time-point. Stacking up, the following Koopman operator relationship $\pmb{\Psi}(Y_f,\Theta) =\mathcal{K} \pmb{\Psi}(Y_p,\Theta)$ should hold.
Therefore, the learning cost function for the deepDMD is given by 
\begin{equation}
\begin{aligned}
\min_{\mathcal{K}, \Theta} & \parallel \pmb{\Psi}(Y_f,\Theta) - \mathcal{K} \pmb{\Psi}(Y_p,\Theta) \parallel_F^2 +  \lambda_1 \parallel \mathcal{K} \parallel_F^2 + \lambda_2 \parallel \Theta \parallel_1
\end{aligned}
\label{eq:deepDMD_cost}
\end{equation}
where $\lambda_1 >0 $ and $\lambda_2 > 0$ indicate the weights for the regularization costs for the Koopman operator and the neural network weights and biases respectively. 
These regularization costs help avoid overfitting of the learned models and additionally, the regularization cost on the Koopman operator promotes robustness. 
We illustrate 
deepDMD in Fig. \ref{fig:deepDMD_overview}. 

Robust DMD \cite{sinha2018robust} is seen as a special case of \eqref{eq:deepDMD_cost} such that $\pmb \Psi(X_t) := X_t$\,, which gives us the following cost function: 
\begin{align}
    \min_{\mathcal{K}} & \parallel Y_f - \mathcal{K} Y_p \parallel_F^2 +  \lambda_1 \parallel \mathcal{K} \parallel_F^2\,.
    \label{eq:RDMD_cost}
\end{align}
Note that for a given value of $\lambda_1$, one can obtain an analytical solution to \eqref{eq:RDMD_cost} such that
\begin{align}
    {\cal K} = Y_f \left(Y_p + \lambda_1 I \right)^{\dagger}
    \label{eq:robust_dmd_solution}
\end{align}
where $I$ is the identity matrix and $(\cdot)^\dagger$ denotes the Moore-Penrose pseudo inverse of $(\cdot)$. 

\stitle{Multi-Step Prediction}. 
The Koopman operator establishes a linear relationship between two consecutive measurement vectors in the space of observables. Repeat application of the Koopman operator on the current measurement ($X_t$) therefore allows us to generate a sequence of predicted measurements over some prediction window $H$ as below: 
\begin{align}
    \pmb\Psi(\widehat{X}_{t+h}|X_t)=\mathcal{K}^h\,\pmb\Psi(X_t)\quad h=1,2,\dots,H\,.
\end{align}
Recall that the predicted measurements $\lbrace\widehat{X}_{t+h}\rbrace_{h=1}^H$ are retrieved trivially from the measurement-inclusive observable $\pmb\Psi$\,.

\section{Graph Neural Network-Based Prediction}
\label{sec-gnn}
We next describe an alternative method that can 
holistically exploit network topology and spatio-temporal information. 
Graph neural networks (GNNs) have been verified to be effective for predictive analysis 
over network data, by jointly considering node and topological features. For this work, we adopt the Spatio-Temporal Graph Convolution Network (STGCN), a form of GNN introduced
in \cite{yu2017spatio} for time-series traffic forecasting, and extend it to support flexible power system transient problem with variable lengths of prediction window (\eg 1-10\,s). Specifically, we develop an enhanced STGCN with an \textit{incremental inference} strategy to make it adaptive to variable testing prediction lengths, while the model is still trained only once. This helps us reduce the re-training effort and make STGCN adaptive to time-series prediction of power system transients.

\stitle{Notations}. We represent the cyber-physical power network as an \textit{attributed} graph, such that $G_t = (V,E,X_t)$ represents a {\em snapshot} of the network at time $t$, with $V$ as the (fixed) set of vertices (or, nodes) representing the buses with sensors (PMU); $E\subseteq V\times V$ as the (fixed) edges between the sensor nodes; and $X_t\in\mathbb{R}^{n}$ as the (time-stamped) $n$-dimensional node feature vector representing the sensor measurements from across the network at time $t$\,. Moreover, $A$ denotes the fixed adjacency matrix representing the connections between the sensor nodes.

Given past $M$ measurements, $\lbrace X_j\rbrace_{j=t-M+1}^t$, of the power network, the problem is to predict the \textit{most likely} measurements in the next $H$ time-steps, $\lbrace X_i\rbrace_{i=t+1}^{t+H}$, as follows:

\begin{align*}
\lbrace \widehat{X}_i\rbrace_{i=t+1}^{t+H} &= 
\operatorname*{argmax}_{\lbrace X_i\rbrace_{i=t+1}^{t+H}} \log P\left(\lbrace X_i\rbrace_{i=t+1}^{t+H} \left| \lbrace X_j\rbrace_{j=t-M+1}^t\right.\right)
\end{align*}

\stitle{Overview of STGCN}.
In a nutshell, an STGCN \cite{yu2017spatio} learns the best representation 
of a power grid with graph convolutional operators 
based on spectral graph convolution. 
A graph convolution operator ``$*_{\mathcal{G}}$'' can be defined as:
\begin{equation}
\label{graph-covl-operator}
g_{\theta} *_{\mathcal{G}} X_t= U g_{\theta}(\Lambda) U^\top X_t\,,
\end{equation}
where $g_{\theta}(\cdot)$ is the graph kernel parameterized by the vector $\theta\in\mathbb{R}^K$ (for some $K$) in the Fourier domain, and $X_t$ is the time-stamped node feature vector (\eg frequency measurements). $U \in \mathbb{R}^{n \times n}$ refers to the graph Fourier basis that is the matrix of eigenvectors of the normalized graph Laplacian $L$ = $I_{n}-D^{-\frac{1}{2}} A D^{-\frac{1}{2}} \in \mathbb{R}^{n \times n}$, 
where $I_{n}$ is the $n$-dimensional identity matrix, and $D\in\mathbb{R}^{n\times n}$ is the diagonal degree matrix with $D_{i i}=\Sigma_{j} A_{i j} \,\forall i$ as derived from the topology of the power network.
$\Lambda \in \mathbb{R}^{n \times n}$ is the diagonal matrix of eigenvalues of the normalized graph Laplacian $L$, and the filter $g_{\theta}(\Lambda)$ is diagonal too. As pointed out in \cite{shuman2013emerging,yu2017spatio}, graph convolution is the multiplication of the graph Fourier transform $U^\top X_t$ of a signal $X_t$ with a filter of graph kernel $g_{\theta}$.

\begin{figure}[h]
    \centering
     \includegraphics[width = 0.7 \columnwidth]{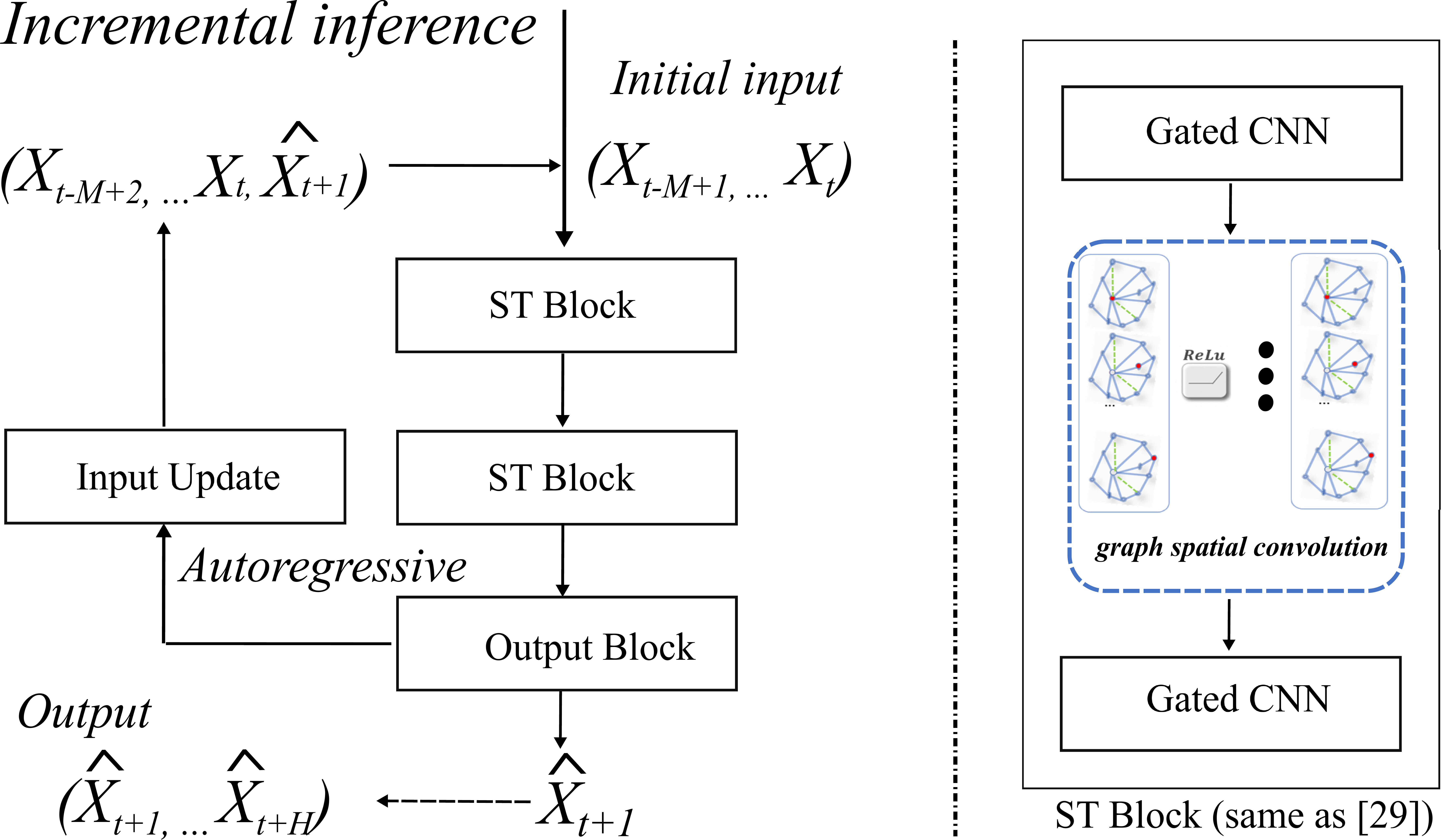}
    \caption{Architecture of our STGCN framework.}
    \label{fig:STGCN_overview}
\end{figure}

\stitle{An Incremental Inference STGCN Model}. 
Fig. \ref{fig:STGCN_overview} shows the architecture of the enhanced STGCN model, developed to support the variable prediction lengths in the power system transient trajectories forecasts. Following the work of \cite{yu2017spatio}, the model consists of spatio-temporal convolutional blocks (ST-blocks), each built by two gated temporal convolution layers and one spatial graph convolution layer in a \textit{sandwich} structure. After stacking  several ST-blocks, an output block integrates the comprehensive features extracted from spatial and temporal domains and outputs a single-step prediction. Extending the work of \cite{yu2017spatio}, our STGCN shuffles the prediction and feeds the output back into the model to generate forecasts with varying prediction length $H$. It concatenates the predictions for the next $H$ time-steps and treats temporal predictions as an autoregressive task. In the following, we briefly outline each of the components.

\etitle{Graph Spatial Convolution}. Existing ML models for power systems time-series prediction (e.g., LSTM \cite{liu2019transient,li2020machine}) do not take into account the underlying spatial structure of the network. The spatial graph convolution layer within the ST-block allows us to extract meaningful patterns and representations in the spatial domain by performing graph convolution directly on graph-structured sensor measurements (e.g., PMUs data). To resolve the high $\mathcal{O}\left(n^{2}\right)$ computation cost of the kernel $g_\theta$, we adopt similar approximation strategies as in \cite{yu2017spatio}. In particular, by approximating $g_{\theta}(\Lambda)\approx\sum_{k=0}^{K-1}\theta_kT_k(\widetilde{\Lambda})$ via Chebyshev polynomials $T_{k}(\cdot)$ of orders up to $(K-1)$, where $\widetilde{\Lambda}=2\Lambda/\lambda_{\max}\!-\!I_n$ and $\lambda_{\max }$ is the largest eigenvalue of $L$, the computation cost can be brought down to $\mathcal{O}(K\,|{E}|)$ \cite{defferrard2016convolutional}. Restricting to $K=2$ (first-order polynomials) and assuming $\lambda_{\max } \approx 2$, $g_{\theta}$ can be further approximated as follows \cite{kipf2016semi}:
\begin{align}
\label{gcn}
    g_{\theta} *_{\mathcal{G}} X_t \approx \theta (\tilde{D}^{-\frac{1}{2}} \tilde{A}\tilde{D}^{-\frac{1}{2}})X_t \,,
\end{align}
where $\theta=\theta_0\!=\!-\theta_1$, $\tilde{A}\!=\!A+I_{n}$ and $\tilde{D}_{i i}\!=\!\sum_{j} \tilde{A}_{i j}$\,, leading to efficient implementation as the matrix product \cite{kipf2016semi}. 

\etitle{Gated Temporal Convolution}
STGCN \cite{yu2017spatio} applies causal convolution \cite{oord2016wavenet} with a width-$K_t$ kernel in time dimension and gated linear unit (GLU) \cite{dauphin2017language} as a non-recurrent approach for fast training (denoted as Gated CNN in Fig. \ref{fig:STGCN_overview}). The input is an $M\times n\times C_i$-dimensional tensor representing a sequence of $M$ graph embeddings (corresponding to the past snapshots) with $C_i$ channels.  
The output with $C_o$ channels is an $(M-K_t+1) \times n \times C_0$-dimensional tensor which shrinks the size by $(K_t-1)$ in the temporal dimension.

\eetitle{Output Block and Model Learning} \cite{yu2017spatio}. One ST-block shrinks the input feature tensor by $2(K_t-\!1)$ in the temporal dimension. 
After stacking two ST-blocks, STGCN attaches an extra gated temporal
convolution layer with a fully-connected layer as the output block. The final output $\widehat{X}_{t+1} \in \mathbb{R}^{n}$ is a measurement prediction (\eg frequency) for all the sensor nodes in the power network, obtained as a minimizer to the following $\mathcal{L}_2$ loss function during training: 
\begin{equation}
L(\widehat{X} ; \Theta)=\sum_{t}\left\|\widehat{X}\left(\{X_{j}\}_{j=t-M+1}^t, \Theta\right)-X_{t+1}\right\|^{2}
\end{equation}
where $\Theta$ refers to the trainable parameters, and $X_{t+1}$ are the available ground truth measurements. 

\stitle{Incremental Inference}. 
As a unique feature, our STGCN model supports an incremental prediction process for real-time power grid prediction. While existing models are trained for a specific predictive window $H$, we developed an adaptive STGCN to support arbitrary prediction length $H'$.  Let us consider, without losing the generality, that the STGCN model is trained based on a prediction window length of $H$, while the inference task for a testing case has a prediction window length of $H'>H$ (the case $H'\leq H$ is trivial). We firstly generate the next $H$ predictions and then shuffle the time-series input with predicted values to construct the new input. Specifically, we augment the observed measurements with the predicted ones $\lbrace \widehat{X}_i\rbrace_{i=t+1}^{t+H}$ and form a new input $\lbrace X_{t+H-M+1}, \ldots, X_t,\widehat{X}_{t+1},\ldots, \widehat{X}_{t+H}\rbrace$ to generate the prediction snapshot $\widehat{X}_{t+H+1}$, and continue until we predict measurements up to $\widehat{X}_{t+H’}$.

\stitle{Hyperparameter Selection}
Following a grid search strategy with 5-fold cross-validation, we select the optimized length of observation window ($M$) to 200 time-points, \ie 4s window at the PMU sampling rate of 50 per second (see Sec.\,\ref{sec:setup}).

\section{Simulation Setup}\label{sec:setup}
\begin{figure}
    \centering
    \includegraphics[width = 0.6 \columnwidth]{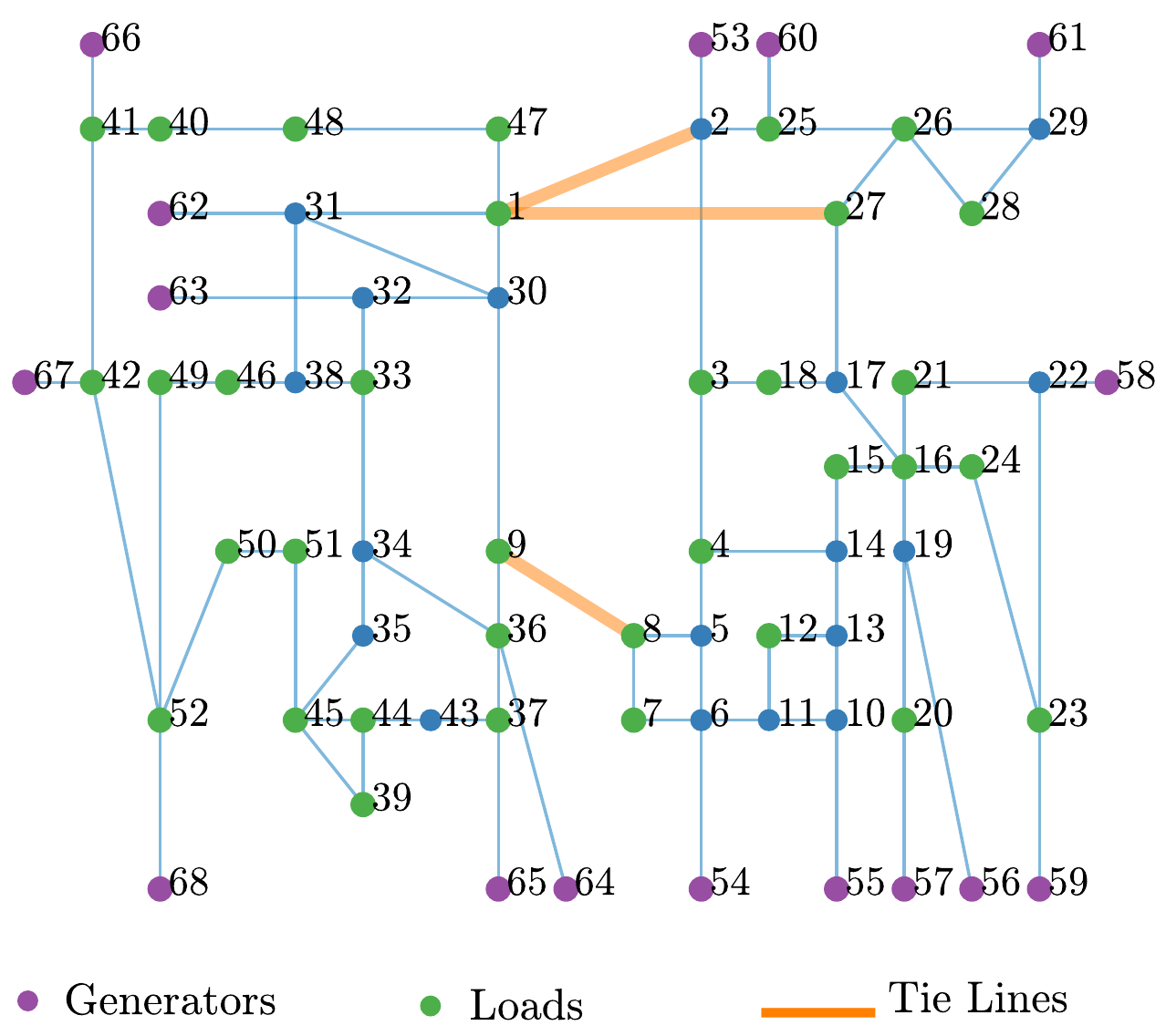}
    \caption{Single-line diagram of IEEE 68-bus system with generator and load locations. Tie lines connecting the two areas are highlighted.}
    \label{fig:IEEE68busSystem}
    \vspace{-3ex}
\end{figure}

The performance of the proposed predictive  models are evaluated in a wide range of scenarios generated using the GridSTAGE simulation framework \cite{gridstage2020}.
We use the IEEE 68-bus system (illustrated in Fig. \ref{fig:IEEE68busSystem}) to 
develop and test the three predictive models: Robust DMD, deepDMD, and STGCN. Data corresponding to the IEEE 68-bus system is generated over various scenarios where each \textit{scenario} corresponds to a unique configuration (e.g., load changes) of the power network. 
Different criteria such as the proximity of the loads to the generators, degree of connectivity of loads, etc, are considered for selecting locations for load changes for creating scenarios (see Table \ref{tab:load_strategies}). 

\stitle{Data Generation}.
\label{subsec:gridstage}
GridSTAGE (Spatio-Temporal Adversarial scenario GEneration)  is a recently developed framework for simulation of adversarial scenarios and generation of multivariate spatio-temporal data in cyber-physical systems \cite{gridstage2020}, and has been used in \cite{nandanoori2020model} to test real-time attack identification algorithm. GridSTAGE is developed based on Matlab, and leverages Power System Toolbox (PST) \cite{chow1992toolbox} where the evolution of power network is governed by nonlinear differential equations that includes synchronous machine models, excitation systems, turbine governors, and automatic generation control.   

Detailed instructions on generating  data scenarios with different system topologies, attack characteristics, load characteristics, sensor configuration, control parameters can be found in 
\cite{gridstage2020}. 
We use GridSTAGE to generate spatio-temporal time-series data with load changes across the network for the IEEE $68$ bus system. 
The PMU data is considered to report high frequency measurements at 50 snapshots per second. In practice, factors such as outdated calibration and inherent communication channel noises lead to inaccuracy in the PMU measurements in the range of 80-90 dB of signal-to-noise ratio (SNR) \cite{bhandari2019real}. To capture this effect, the synthetic PMU measurements from GridSTAGE have been corrupted by noise with an SNR value of 85 dB.

\eetitle{Scenarios generation}. Latin hypercube sampling (LHS) is applied to generate multiple scenarios representing different combinations of changes in the active power load magnitudes across the network. {Under every scenario, at each load bus, the load changes are implemented as a pulse where the pulse width is chosen to be 0.25 seconds, that is, the load magnitude is either increased or decreased, held there for 0.25 seconds and reverted. This time instance is treated as time} $t=0$ {for collecting the time-series data. All the loads were changed at the same time and the magnitude of load change is different at each load bus which is chosen based on the LHS. }
LHS is a statistical method that allows us to generate random samples of load changes (deviations from nominal) from multivariate distributions. We group the load change magnitudes broadly into three classes: \textit{low}, \textit{medium}, and \textit{high}, to generate a wide range of scenarios representative of different loading conditions for the final simulations. 
The transient simulations data for the power network are generated and recorded for training and testing from over 330 different scenarios summarized in Table \ref{tab:load_strategies}. Out of those, $30$ scenarios are chosen for training and validation of the individual models (Robust DMD, deepDMD, and STGCN), while the remaining 300 are used for rigorous testing and performance evaluation. In the scope of this study, the Koopman and STGCN models are trained assuming PMUs are located at each bus. However, if PMUs are located only at a subset of buses in the system, then the models trained on the available PMUs enables prediction only at those corresponding buses. 

\begin{table}
\begin{center}
\caption{A summarized list of the different load change scenarios simulated for generating extensive datasets for training and testing of the three predictive models: \textit{Robust DMD}, \textit{deepDMD} and \textit{STGCN}. Three SD values are chosen to represent different magnitudes of load changes: low, medium and high. Various strategic locations for load changes are selected to generate 330 scenarios out of which 30 scenarios were used for training, validation and the remaining 300 for testing. 
To quantify the prediction performance, RMSE and MAPE errors are chosen and they are computed by taking the average of these errors over 20 testing scenarios for a given choice of SD value. Color coding: cells with the lowest error values are shown in green; cells with values around $50^{th}$ percentile are shown in yellow; while cells with the highest values are shown in red.}
\label{tab:load_strategies}
\resizebox{\textwidth}{!}{  
\begin{tabular}{ccc|c|c|c|c|c|c|}
\cline{4-9}
                            &                                      & \multicolumn{1}{l|}{}      & \multicolumn{2}{c|}{Robust DMD}                                                                                                                & \multicolumn{2}{c|}{deepDMD}                                                                                                                   & \multicolumn{2}{c|}{STGCN}                                                                                                                     \\ \hline
\multicolumn{1}{|c|}{\begin{tabular}[c]{@{}c@{}}Case \\ Index\end{tabular}    }   & \multicolumn{1}{c|}{\begin{tabular}[c]{@{}c@{}}Load Change \\ Location\end{tabular}    }        & {\begin{tabular}[c]{@{}c@{}}Load Change \\ Magnitude\end{tabular}    }                          & \begin{tabular}[c]{@{}c@{}}RMSE {[}Hz{]}\\ ($\times 10^{-3}$)\end{tabular} & \begin{tabular}[c]{@{}c@{}}MAPE\\ ($\times 10^{-5}$)\end{tabular} & \begin{tabular}[c]{@{}c@{}}RMSE {[}Hz{]}\\ ($\times 10^{-3}$)\end{tabular} & \begin{tabular}[c]{@{}c@{}}MAPE\\ ($\times 10^{-5}$)\end{tabular} & \begin{tabular}[c]{@{}c@{}}RMSE {[}Hz{]}\\ ($\times 10^{-3}$)\end{tabular} & \begin{tabular}[c]{@{}c@{}}MAPE\\ ($\times 10^{-5}$)\end{tabular} \\ \hline
\multicolumn{1}{|c|}{}      & \multicolumn{1}{c|}{Loads 1-hop}     & low  & \cellcolor[HTML]{C8D666}{\color[HTML]{000000} 6.51}                        & \cellcolor[HTML]{D4DB6D}{\color[HTML]{000000} 7.01}               & \cellcolor[HTML]{FFCD71}{\color[HTML]{000000} 11.28}                       & \cellcolor[HTML]{FEB461}{\color[HTML]{000000} 13.34}              & \cellcolor[HTML]{E2DF74}{\color[HTML]{333333} 7.57}                        & \cellcolor[HTML]{FFEA84}{\color[HTML]{333333} 8.84}               \\ \cline{3-9} 
\multicolumn{1}{|c|}{1}     & \multicolumn{1}{c|}{away from}       & medium  & \cellcolor[HTML]{FFE380}{\color[HTML]{000000} 9.4}                         & \cellcolor[HTML]{FFDE7D}{\color[HTML]{000000} 9.79}               & \cellcolor[HTML]{FFCE72}{\color[HTML]{000000} 11.2}                        & \cellcolor[HTML]{FEBB65}{\color[HTML]{000000} 12.78}              & \cellcolor[HTML]{FFD677}{\color[HTML]{333333} 10.47}                       & \cellcolor[HTML]{FEC26A}{\color[HTML]{333333} 12.16}              \\ \cline{3-9} 
\multicolumn{1}{|c|}{}      & \multicolumn{1}{c|}{generators}      & high & \cellcolor[HTML]{FFD375}{\color[HTML]{000000} 10.73}                       & \cellcolor[HTML]{FFCD71}{\color[HTML]{000000} 11.26}              & \cellcolor[HTML]{FC7A3B}{\color[HTML]{000000} 18.25}                       & \cellcolor[HTML]{FB5725}{\color[HTML]{000000} 21.16}              & \cellcolor[HTML]{FDA658}{\color[HTML]{333333} 14.56}                       & \cellcolor[HTML]{FD8945}{\color[HTML]{333333} 16.96}              \\ \hline
\multicolumn{1}{|c|}{}      & \multicolumn{1}{c|}{Loads 2-hops}    & low  & \cellcolor[HTML]{AECD58}{\color[HTML]{000000} 5.42}                        & \cellcolor[HTML]{BDD260}{\color[HTML]{000000} 6.06}               & \cellcolor[HTML]{FFE17E}{\color[HTML]{000000} 9.6}                         & \cellcolor[HTML]{FEC86E}{\color[HTML]{000000} 11.64}              & \cellcolor[HTML]{C2D463}{\color[HTML]{333333} 6.28}                        & \cellcolor[HTML]{F9E881}{\color[HTML]{333333} 8.54}               \\ \cline{3-9} 
\multicolumn{1}{|c|}{2}     & \multicolumn{1}{c|}{away from}       & medium  & \cellcolor[HTML]{C2D463}{\color[HTML]{000000} 6.28}                        & \cellcolor[HTML]{DDDE71}{\color[HTML]{000000} 7.37}               & \cellcolor[HTML]{FFCC70}{\color[HTML]{000000} 11.35}                       & \cellcolor[HTML]{FEBB65}{\color[HTML]{000000} 12.79}              & \cellcolor[HTML]{E1DF74}{\color[HTML]{333333} 7.56}                        & \cellcolor[HTML]{FFDB7B}{\color[HTML]{333333} 10.03}              \\ \cline{3-9} 
\multicolumn{1}{|c|}{}      & \multicolumn{1}{c|}{generator buses} & high & \cellcolor[HTML]{E3E075}{\color[HTML]{000000} 7.64}                        & \cellcolor[HTML]{FFEA84}{\color[HTML]{000000} 8.79}               & \cellcolor[HTML]{FEB863}{\color[HTML]{000000} 13.05}                       & \cellcolor[HTML]{FD9E53}{\color[HTML]{000000} 15.17}              & \cellcolor[HTML]{FEC16A}{\color[HTML]{333333} 12.25}                       & \cellcolor[HTML]{FD8C47}{\color[HTML]{333333} 16.73}              \\ \hline
\multicolumn{1}{|c|}{}      & \multicolumn{1}{c|}{Loads 3-hops}    & low  & \cellcolor[HTML]{A6CA54}{\color[HTML]{000000} 5.12}                        & \cellcolor[HTML]{B7D05D}{\color[HTML]{000000} 5.8}                & \cellcolor[HTML]{C5D565}{\color[HTML]{000000} 6.38}                        & \cellcolor[HTML]{EAE279}{\color[HTML]{000000} 7.92}               & \cellcolor[HTML]{6EB736}{\color[HTML]{333333} 2.82}                        & \cellcolor[HTML]{7DBC3E}{\color[HTML]{333333} 3.43}               \\ \cline{3-9} 
\multicolumn{1}{|c|}{3}     & \multicolumn{1}{c|}{away from}       & medium  & \cellcolor[HTML]{B3CF5B}{\color[HTML]{000000} 5.65}                        & \cellcolor[HTML]{C6D665}{\color[HTML]{000000} 6.42}               & \cellcolor[HTML]{EFE47B}{\color[HTML]{000000} 8.14}                        & \cellcolor[HTML]{FFDD7C}{\color[HTML]{000000} 9.9}                & \cellcolor[HTML]{80BD40}{\color[HTML]{333333} 3.54}                        & \cellcolor[HTML]{8DC147}{\color[HTML]{333333} 4.06}               \\ \cline{3-9} 
\multicolumn{1}{|c|}{}      & \multicolumn{1}{c|}{generator buses} & high & \cellcolor[HTML]{BDD260}{\color[HTML]{000000} 6.04}                        & \cellcolor[HTML]{D5DB6D}{\color[HTML]{000000} 7.04}               & \cellcolor[HTML]{FEE983}{\color[HTML]{000000} 8.73}                        & \cellcolor[HTML]{FFD476}{\color[HTML]{000000} 10.63}              & \cellcolor[HTML]{96C54C}{\color[HTML]{333333} 4.45}                        & \cellcolor[HTML]{A8CB55}{\color[HTML]{333333} 5.21}               \\ \hline
\multicolumn{1}{|c|}{}      & \multicolumn{1}{c|}{Loads with}      & low  & \cellcolor[HTML]{9AC64E}{\color[HTML]{000000} 4.6}                         & \cellcolor[HTML]{95C44B}{\color[HTML]{000000} 4.39}               & \cellcolor[HTML]{C1D463}{\color[HTML]{000000} 6.24}                        & \cellcolor[HTML]{DDDE72}{\color[HTML]{000000} 7.39}               & \cellcolor[HTML]{46A921}{\color[HTML]{333333} 1.13}                        & \cellcolor[HTML]{4DAB25}{\color[HTML]{333333} 1.45}               \\ \cline{3-9} 
\multicolumn{1}{|c|}{4}     & \multicolumn{1}{c|}{high degree of}  & medium  & \cellcolor[HTML]{ACCD58}{\color[HTML]{000000} 5.37}                        & \cellcolor[HTML]{BBD25F}{\color[HTML]{000000} 5.98}               & \cellcolor[HTML]{DEDE72}{\color[HTML]{000000} 7.44}                        & \cellcolor[HTML]{FFE480}{\color[HTML]{000000} 9.29}               & \cellcolor[HTML]{65B431}{\color[HTML]{333333} 2.43}                        & \cellcolor[HTML]{70B737}{\color[HTML]{333333} 2.89}               \\ \cline{3-9} 
\multicolumn{1}{|c|}{}      & \multicolumn{1}{c|}{connectivity} & high & \cellcolor[HTML]{B0CE59}{\color[HTML]{000000} 5.51}                        & \cellcolor[HTML]{BFD361}{\color[HTML]{000000} 6.13}               & \cellcolor[HTML]{F3E57D}{\color[HTML]{000000} 8.28}                        & \cellcolor[HTML]{FFD778}{\color[HTML]{000000} 10.43}              & \cellcolor[HTML]{67B432}{\color[HTML]{333333} 2.51}                        & \cellcolor[HTML]{70B737}{\color[HTML]{333333} 2.87}               \\ \hline
\multicolumn{1}{|c|}{}      & \multicolumn{1}{c|}{Loads with}      & low  & \cellcolor[HTML]{C7D666}{\color[HTML]{000000} 6.49}                        & \cellcolor[HTML]{E0DF73}{\color[HTML]{000000} 7.5}                & \cellcolor[HTML]{FFCA6F}{\color[HTML]{000000} 11.51}                       & \cellcolor[HTML]{FEAE5D}{\color[HTML]{000000} 13.88}              & \cellcolor[HTML]{FFD375}{\color[HTML]{333333} 10.71}                       & \cellcolor[HTML]{FEAD5D}{\color[HTML]{333333} 13.92}              \\ \cline{3-9} 
\multicolumn{1}{|c|}{5}     & \multicolumn{1}{c|}{low degree of}   & medium  & \cellcolor[HTML]{FFDE7C}{\color[HTML]{000000} 9.81}                        & \cellcolor[HTML]{FFC96F}{\color[HTML]{000000} 11.57}              & \cellcolor[HTML]{FD8945}{\color[HTML]{000000} 16.96}                       & \cellcolor[HTML]{FC7336}{\color[HTML]{000000} 18.86}              & \cellcolor[HTML]{FD8F49}{\color[HTML]{333333} 16.48}                       & \cellcolor[HTML]{FB5020}{\color[HTML]{333333} 21.78}              \\ \cline{3-9} 
\multicolumn{1}{|c|}{}      & \multicolumn{1}{c|}{connectivity} & high & \cellcolor[HTML]{FFCA6F}{\color[HTML]{000000} 11.48}                       & \cellcolor[HTML]{FEAF5E}{\color[HTML]{000000} 13.76}              & \cellcolor[HTML]{FC7035}{\color[HTML]{000000} 19.09}                       & \cellcolor[HTML]{FB5D29}{\color[HTML]{000000} 20.67}              & \cellcolor[HTML]{FC6C32}{\color[HTML]{333333} 19.38}                       & \cellcolor[HTML]{F91F00}{\color[HTML]{333333} 25.85}              \\ \hline
\end{tabular}}
\end{center}
\end{table}

\stitle{Model Implementation and Configuration}. 
All the models are implemented with TensorFlow and the respective codes are available online \cite{gridprediction2021}.  
The KOT-based predictive models (Robust DMD and deepDMD) are trained and tested on the servers with 2.6 GHz Intel core i7 processor and 16GB of RAM. For deepDMD, we found neural networks with 4 layers can identify the observables in the latent space (for the IEEE 68 bus system). Robust DMD and deepDMD are trained with the regularization weights $\lambda_1 = 0.01$, $\lambda_2 = 0.005$, respectively. All of these KOT-based models use both frequency and voltage magnitude states as training data. Therefore, \textit{these methods yield frequency as well as voltage magnitude predictions}.
On the other hand, the STGCN model is optimized on Google CoLab service with a machine of 4 CPU cores (Intel Xeon @ 2.20 GHz), 25 GB RAM, and an NVIDIA P100 GPU. We use only the frequency state to train STGCN model for frequency prediction, following the setting in~\cite{yu2017spatio} that assumes the training and testing data 
refer to the same measurement.  

\stitle{Performance Metrics}. For convenience and uniformity across scenarios, we will denote the current time as $t=0$ and consider the predicted measurements $\lbrace \widehat{X}_t\rbrace_{t=1}^H$ at the next $H$ time-steps $t=1,2,\dots,H$\,. We use the following error metrics for quantifying and comparing the prediction accuracy of the proposed models: \textit{root mean square error} (RMSE) 
and \textit{mean absolute percentage error} (MAPE), as defined below for each sensor location (bus-$i$): 
\begin{align*}
    \text{RMSE}_i &:= \sqrt{\frac{1}{H} \sum_{t=1}^H(X_{t,i} - \widehat{X}_{t,i})^2}\\
    \text{MAPE}_i &:= \frac{1}{H} \sum_{t=1}^H \left|\frac{X_{t,i} - \widehat{X}_{t,i}}{X_{t,i}}\right|\,,
\end{align*}
where $X_{t,i}$ and $\widehat{X}_{t,i}$ respectively denote the actual (\textit{ground truth}) and predicted values of frequency at location bus-$i$ at time $t$\,. 
As the objective is to predict the frequencies acrosss the IEEE 68-bus network, the RMSE$_i$ and MAPE$_i$ are defined for each bus-$i$. For the purpose of reporting these errors in Table \ref{tab:load_strategies}, however, the worst case (or, maximum) values of RMSE and MAPE across all the buses, \ie $\max_i \text{RMSE}_i$ and $\max_i \text{MAPE}_i$\,, are considered.

\section{Performance Evaluation}\label{sec:results}
In this section, we present a detailed analysis of the performances of the different predictive models based on their \textit{overall, temporal and spatial} accuracy, sensitivities to model parameters and time costs.

\begin{figure*}
    \centering
    \includegraphics[width = 0.99 \textwidth]{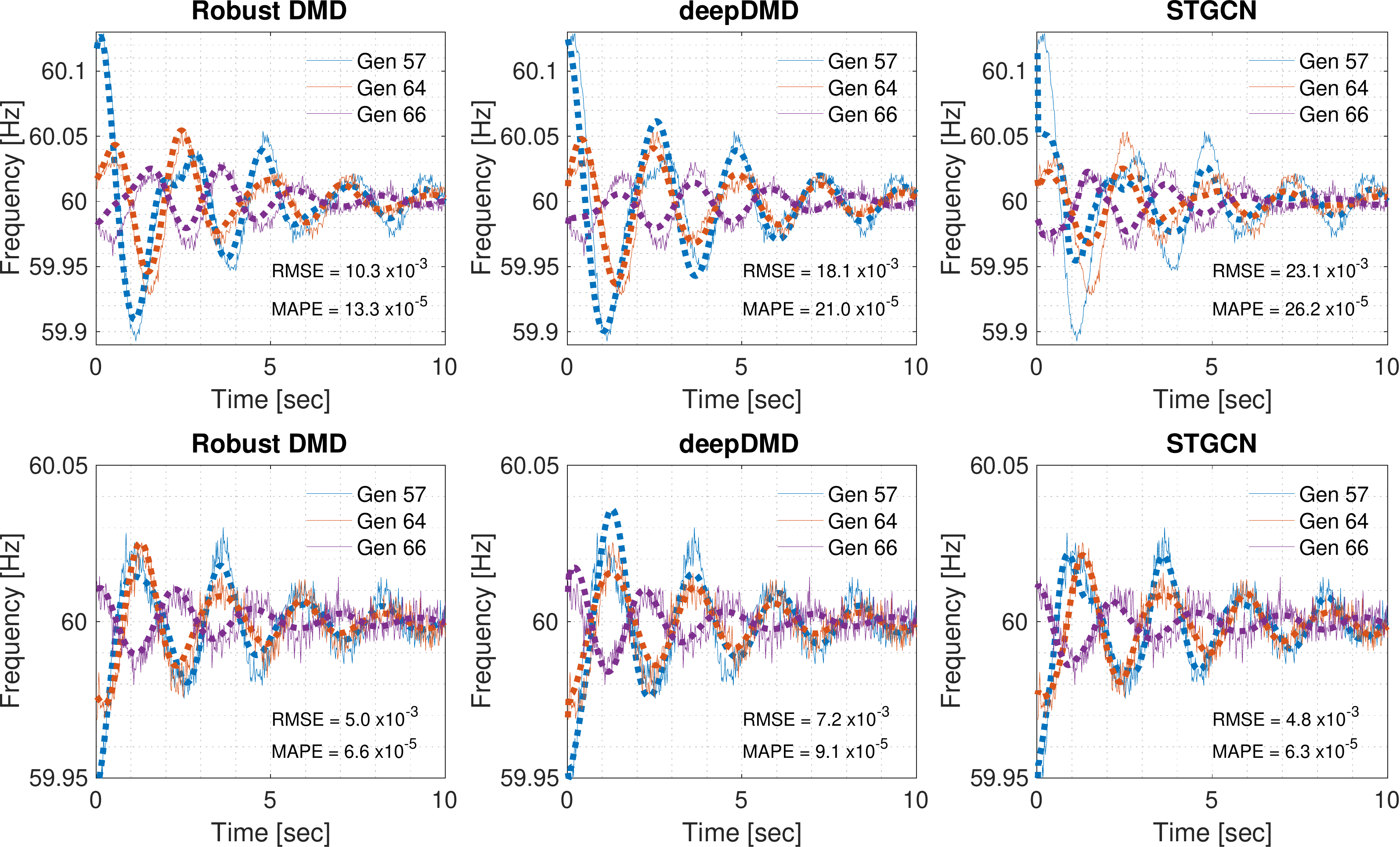}
    \caption{Actual and predicted frequencies corresponding to two randomly chosen disturbances using Robust DMD, deepDMD and STGCN at generator buses 57, 64, 66 for a period of 10 seconds (500 time-steps as per the given PMU sampling frequency.) The top plots correspond to disturbance\,-\,1 (load changes at buses 20, 23, 25, 36, 41, 42) and the bottom plots correspond to disturbance\,-\,2 (load changes at buses 4, 8, 12, 15, 18, 27, 44, 46, 47, 48, 51). The solid lines show the actual noisy  measurements whereas the dotted lines indicate the predictions.}  
    \label{fig:actual_predicted_comparisons}
\end{figure*}

\stitle{Illustrative Examples}. 
Before presenting the detailed performance analysis, let us illustrate the working of the predictive models on a couple of load change scenarios. Specifically, using measurements corrupted by noise as the initial observation, we verify the accuracy of the predictive models by investigating how well they can ``recover'' the actual (\textit{ground truth}) frequencies. Figure \ref{fig:actual_predicted_comparisons} illustrates the actual and predicted frequencies at 3 arbitrarily chosen generator buses under two different disturbance scenarios: 1) \textit{(top)} disturbance\,-\,1 correspond to the load changes at buses 20,\,23,\,25,\,36,\,41,\,42; and 2) \textit{(bottom)} disturbance\,-\,2 correspond to the load changes at buses 4,\,8,\,12,\,15,\,18,\,27,\,44,\,46,\,47,\,48,\,51. Notice that during the transient period the model predictions demonstrate some undershoot -- \eg Robust DMD, deepDMD for disturbance-2 from 4s and STGCN for disturbance-1 throughout the prediction window -- or overshoot behavior -- \eg deepDMD predictions under disturbance-1 and 2 around 2s.
 
{In the STGCN prediction for disturbance-1, we notice from Fig.} \ref{fig:actual_predicted_comparisons} {that the initial predicted state of the generators change abruptly. Moreover, this behavior is not seen in the disturbance-2.} 
\begin{remark}
{Based on our evaluation study on the different scenarios, the abrupt jump in the predictions (immediately when the prediction is started) is only seen with the STGCN model when the initial disturbance is relatively high (similar to the disturbance-1 scenario in Fig. } \ref{fig:actual_predicted_comparisons}). 
\label{rm:obsv_on_STGCN}
\end{remark}
Overall, all the models successfully ``recover'' the post-disturbance transient evolution of the power network frequencies, with very small RMSE ($<21 \times 10^{-3}\,$Hz) and MAPE ($<23\times 10^{-5}$) values at all locations at all time-points.

\stitle{Overall Accuracy}. 
Next, we report the \textit{overall} accuracy of these predictive models under various cases listed in Table \ref{tab:load_strategies}. Each of the five cases refers to a selection strategy for the load change locations, while under each case three different choices have been considered to represent different magnitudes of load changes: low, medium and high. For each of the 15 combinations of the load change location and magnitude, a set of 22 scenarios have been generated. Out of those 330 scenarios, 30 (2 from each of the 15 combinations of load change location and magnitude) were used for training and validation, while the rest 300 were used for extensive testing. In general, all models can maintain good accuracy, \eg RMSE values within $1.13\times10^{-3}\,$Hz and $19.38\times10^{-3}\,$Hz and MAPE values within $1.45\times10^{-5}$ and $25.85\times10^{-5}$. 
{Among all the cases studied, the best accuracy is achieved with the STGCN in case 4 with low load magnitude changes (with $1.13\times 10^{-3}\,$Hz RMSE and $1.45\times 10^{-5}$ MAPE),  
possibly attributed to its exploitation of spatio-temporal and topological features embedded in the power network. On the flip side, though, the performance of the STGCN was found to be more sensitive to the changes of spatial and network topological features such as degree of connectivity of the load change location (}\eg {cases 4 and 5), in contrast to Robust DMD which displays a much smaller variance in RMSE and MAPE across all the scenarios.}

Furthermore, we make the following general observations for all the predictive models: 
\begin{enumerate}[leftmargin=0.5cm]
    \item 
    when the magnitude of load disturbance is increased, the prediction errors of all the models are nearly doubled;
    \item higher predictive errors are 
    observed when the load disturbance is near a generator bus (\ie \textit{1-hop}) rather than  farther away from one (\eg \textit{2- or 3-hops}); and
    \item prediction errors are small when the load change is at buses with high degree of connectivity (4-5 neighbors) as opposed to low degree of connectivity (1-2 neighbors).
\end{enumerate}
The results summarized above (and in Table\,\ref{tab:load_strategies}) demonstrate the overall performance and comparative analysis of the three predictive models in generating transient trajectory forecasts robustly across various scenarios. Next we further explore the robustness of the performance on \textit{temporal} and \textit{spatial} scales.

\begin{figure}[h!]
    \centering
    \includegraphics[width = 0.6\columnwidth]{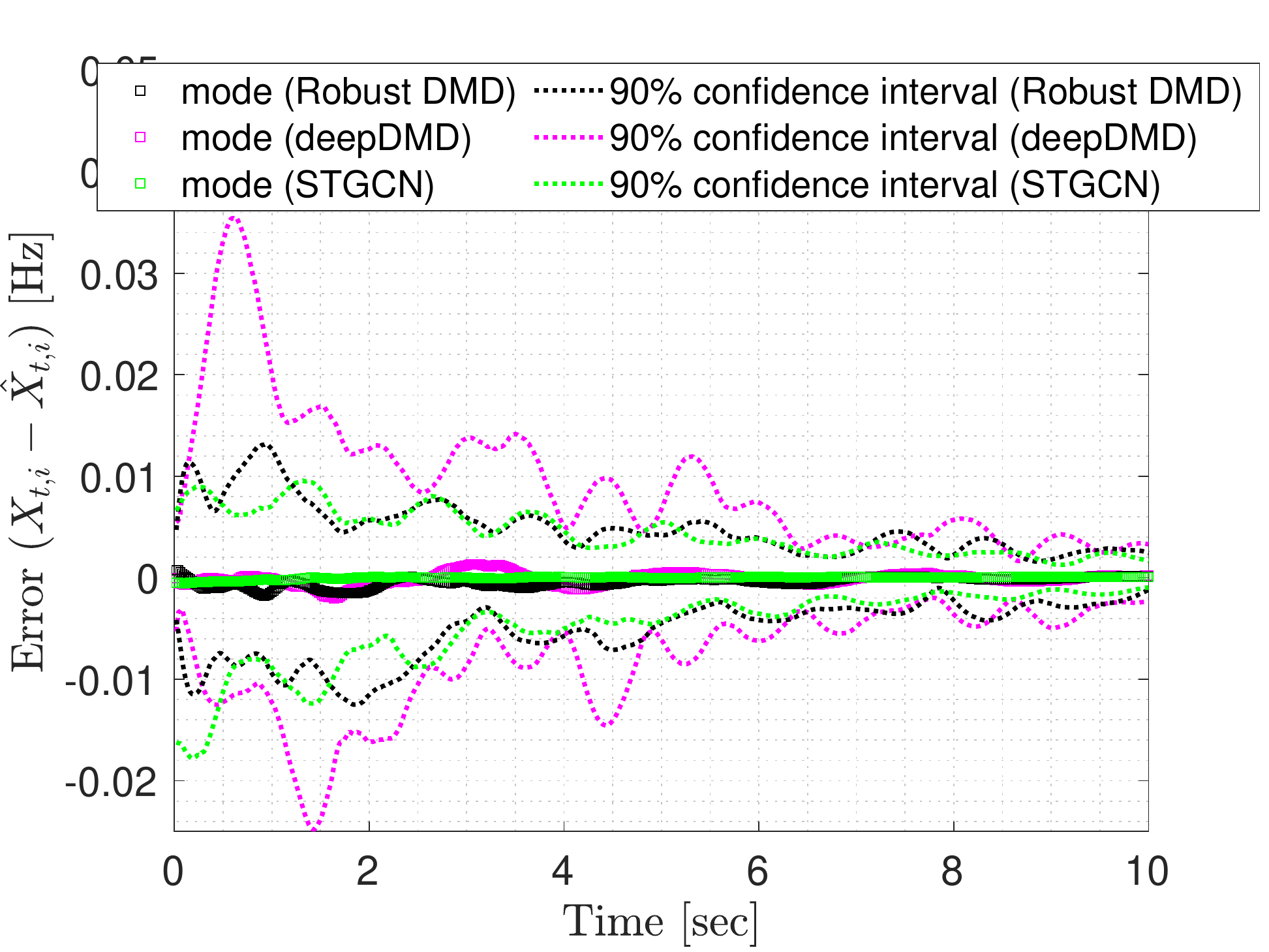}
    \caption{Prediction error of frequencies with $90\%$ confidence interval under several testing scenarios using Robust DMD, deepDMD and STGCN at every time-point.}
    \label{fig:temporal_error}
    \vspace{-2ex}
\end{figure}

\stitle{Temporal Accuracy}. 
To evaluate the model performance on the temporal scale, we take a closer look at how the prediction error evolves as a function of time over a prediction window of 10s. In order to do, we collect the prediction errors $(X_{t,i}\!-\!\widehat{X}_{t,i})$ across all scenarios, and report two statistical measures -- the \textit{mode} (\ie the most frequently occurred value) and a corresponding \textit{90\% confidence interval} -- of the errors at each time-point within the prediction window. Figure\,\ref{fig:temporal_error} shows the modes and the corresponding 90\% confidence intervals of the prediction error values at each time-point within the prediction window. Notice that the modes of the prediction error are nearly zero for all the models uniformly across the prediction window. Moreover, the confidence interval reveals relatively wider distribution of error values during the initial transients period, \ie immediately after a disturbance, before gradually tapering off. Interestingly, the error distributions are larger for the deepDMD method, as compared to Robust DMD and STGCN. Overall, all the predictive models capture the transient evolution of the system frequencies well. 

\begin{figure}[h!]
    \centering
    \includegraphics[width = 0.6\columnwidth]{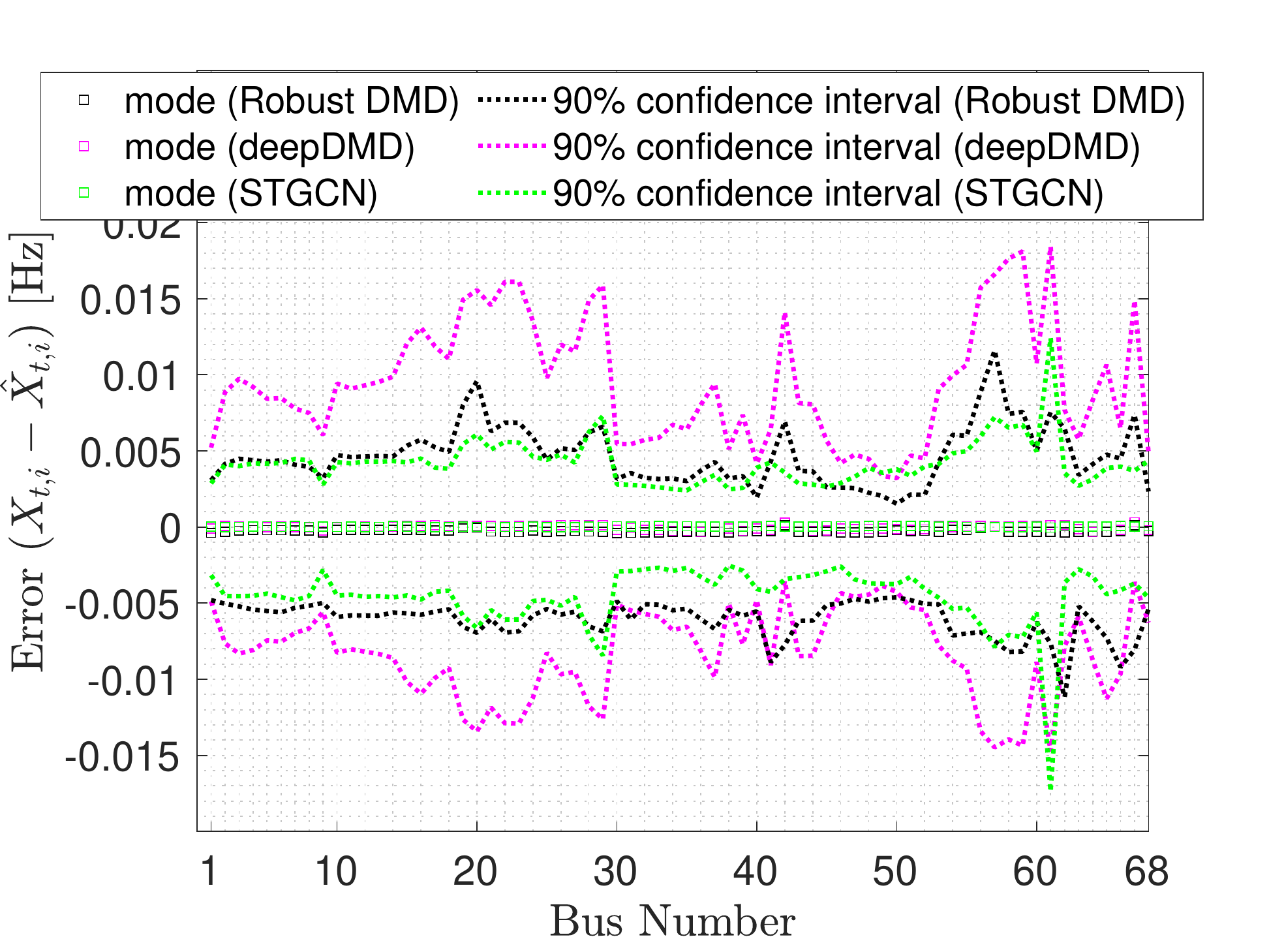}
    \caption{Prediction error of frequencies with 
     $90\%$ confidence interval under several testing scenarios using Robust DMD, deepDMD and STGCN; mode error aggregated over all buses.}
    \label{fig:spatial_error}
    \vspace{-2ex}
\end{figure}

\stitle{Spatial Accuracy}.
We next evaluate the model performance on the spatial scale, by evaluating the prediction accuracy at each bus on the network. Our goal is to understand the impact of bus locations in the network on the transient prediction performance. For all the test scenarios, the prediction errors ($X_{t,i}-\widehat{X}_{t,i}$) are aggregated at each bus as averaged value over all time-points, as reported in Fig. \ref{fig:spatial_error}. 
We observe that the modes of the prediction errors with respect to each predictive model remain close to zero at every bus. We notice some non-uniformity in how the 90\% confidence interval is spatially distributed across the network, \eg the predictive errors display a relatively wider range of values for load buses 18-to-30 and generator buses 53-to-63. Interestingly, though, the error distribution (\ie the 90\% confidence intervals) appears to follow the same pattern for each of the predictive models, indicating that the spatial error distribution perhaps reveals an inherent characteristic of power system network under study. Finally, while all the predictive models demonstrate robust performance on the spatial scale across all scenarios, the error distributions of deepDMD are generally larger than those of Robust DMD and STGCN.

\eetitle{Robustness}. In summary, the results presented above demonstrate the robustness of the performance of the predictive models from the following two perspectives:
\begin{enumerate}[leftmargin=0.45cm]
    \item all the predictive models limit the temporal and spatial errors in the range of $[-0.025, 0.04]$ Hz (\textit{90\% confidence intervals}), as per Figs.\,\ref{fig:temporal_error}-\ref{fig:spatial_error}, across all scenarios in Table\,\ref{tab:load_strategies}. 
    \item all the test studies for prediction performance were carried out under sensor measurement noise of 85\,dB. 
\end{enumerate}

\begin{figure*}[h]
    \centering
    \includegraphics[width = 0.8\textwidth]{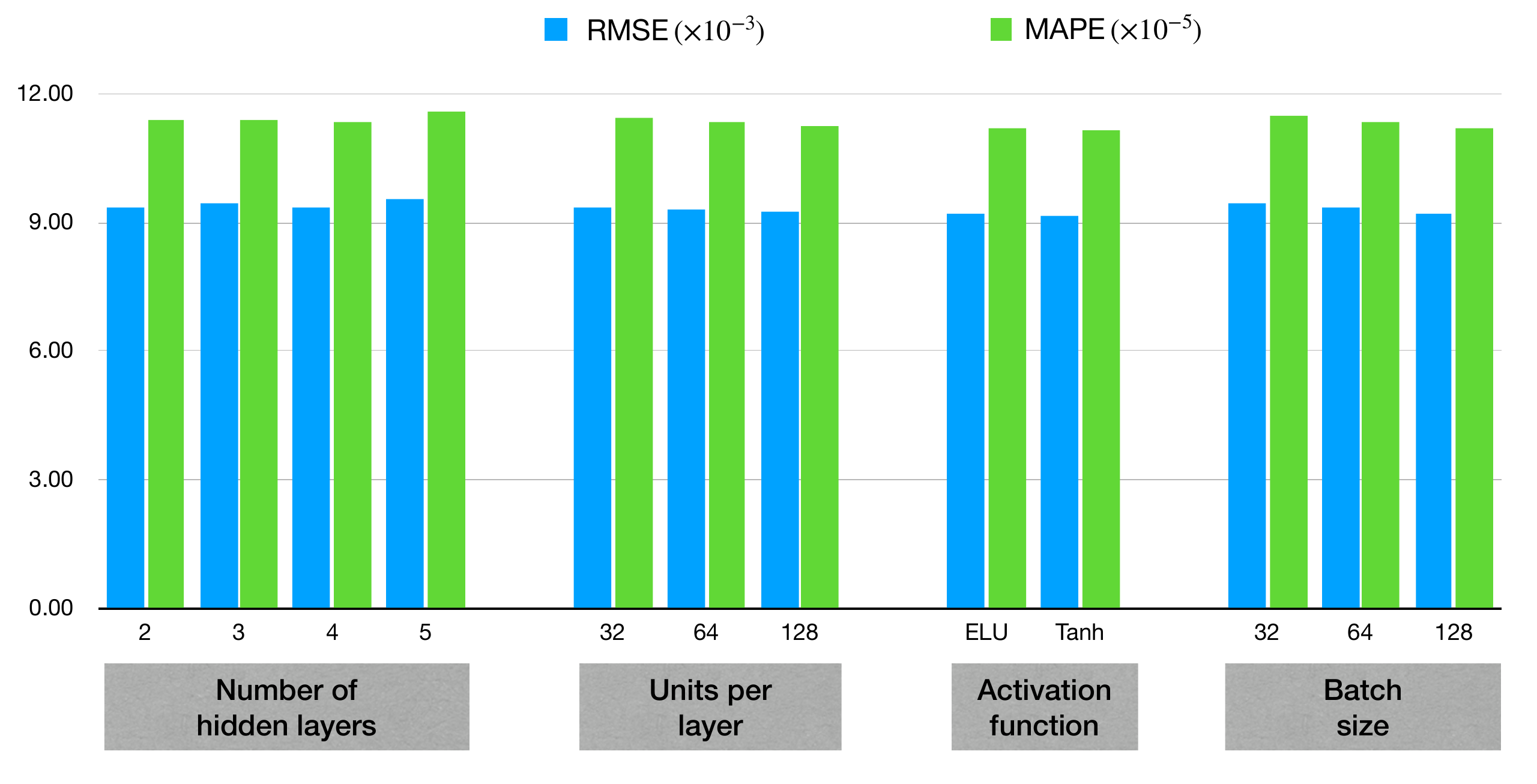}
    \caption{RMSE for several choice of hyperparameters for case-4 with SD = 13. Similar pattern is seen over other cases as well.}
    \label{fig:error_bar_hyper}
    \vspace{-2ex}
\end{figure*}

\stitle{Sensitivity to Model Parameters}.
The deepDMD model is seen to provide reliable predictions over a range of choice of hyperparameters, as shown in Fig. \ref{fig:error_bar_hyper}, with respect to the number of hidden layers, number of units per layer, activation functions and the batch size. This demonstrates that the deepDMD is robust to the choice of hyperparameters.  

\begin{figure}[h!]
    \centering
    \includegraphics[width = 0.6\columnwidth]{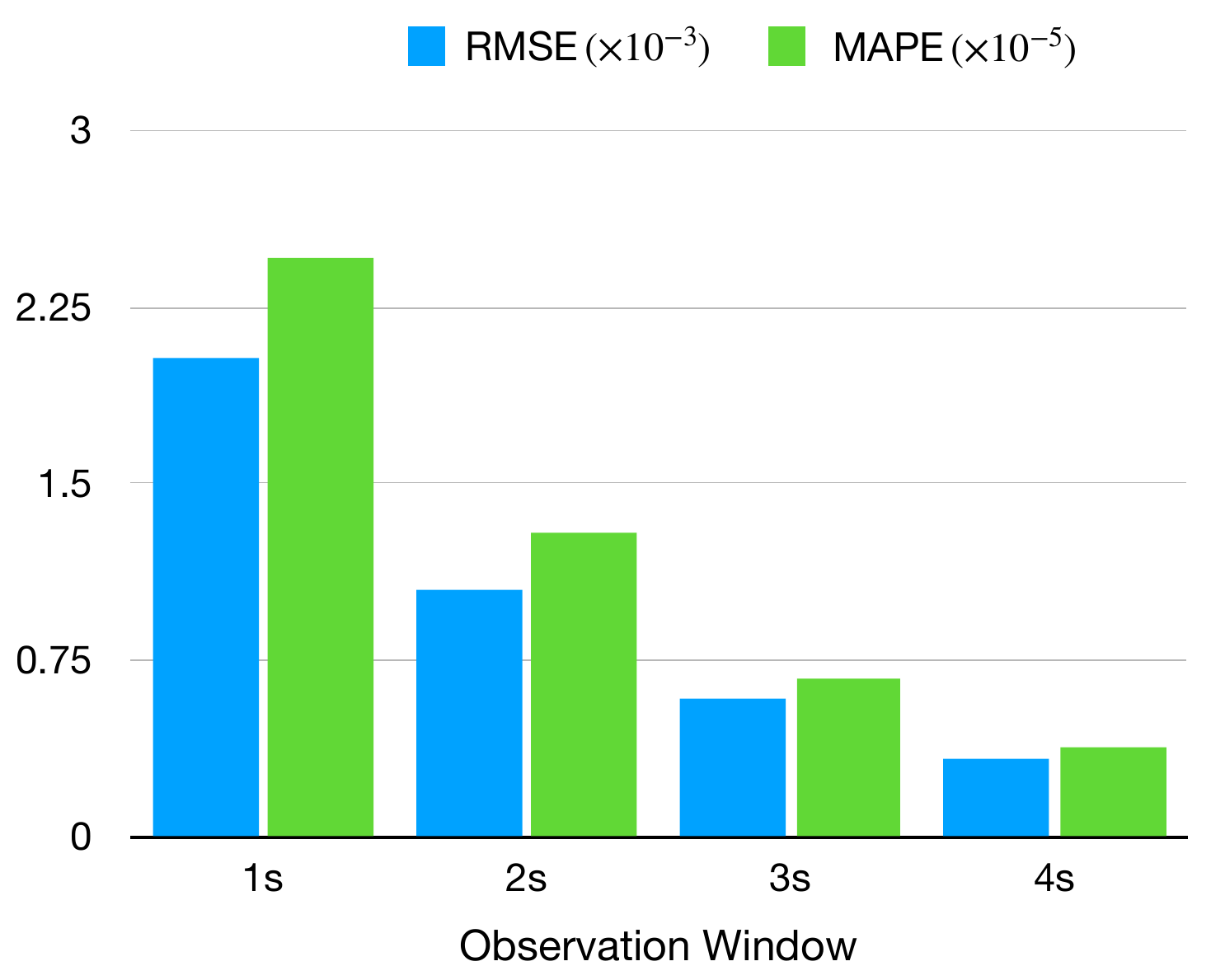}
    \caption{STGCN prediction performance (for a period of 10s) as a function of observation window after the disturbance.}
    \label{fig:STGCN_error_history}
    \vspace{-2ex}
\end{figure}

STGCN-based predictive model requires a history of observations for predicting future state evolution. 
Figure \ref{fig:STGCN_error_history} shows the effect of observation length (1s, 2s, 3s, 4s) on the accuracy of the predictions using STGCN. Perhaps as expected, longer observation window leads to better prediction results. Specifically, as the length of the observation window increases from 1s to 4s, STGCN records 83.7\% improvement in RMSE. Note that a 4s observation window was used for all STGCN prediction results reported above.

\stitle{Time Cost}. 
Given the same training environment, the training of deepDMD takes on average 10-60 minutes depending on the choice of hyperparameters. Since, Robust DMD admits an analytical solution as shown in \eqref{eq:robust_dmd_solution}, training time is only around 5s. In CoLab environment, STGCN takes on average  50 minutes to train for the same amount of training data. All models are quite feasible for fast prediction. For a prediction window of 10s (\ie 500 data points at 50 PMU samples per second), both robust DMD and deepDMD has an average inference cost of 0.1s, while that of STGCN ranges from 1.2s (with 1s observation window) to 2.1s (with 4s observation window). Note the trade-off between the inference time cost and the prediction RMSE in STGCN performance with varying observation window length.

\stitle{Summary}. 
In general, all the predictive models demonstrate robust performance under noisy measurements with high accuracy (RMSE within $19.38\times10^{-3}$\,Hz and MAPE within $25.85\times10^{-5}$), on both \textit{temporal} and \textit{spatial} scales, across all of the 300 load change scenarios listed in Table\,\ref{tab:load_strategies}, alongside demonstrating feasibility of practical implementation with fast inference (0.1-2.1s for a 10s prediction window). 
In particular, KOT-based methods (Robust DMD and deepDMD) model the power system evolution as a linear dynamical system (in \textit{lifted} space), do not require the knowledge of the network topology, and deliver very fast inference (0.1s for 10s prediction window). STGCN, on the other hand, achieves best accuracy by taking advantage of enriched spatio-temporal features in (and is thus more sensitive to) available historical data and network topology, and, with the proposed enhancement on \textit{incremental inferencing}, easily adapts to variable prediction window length, much like the KOT-based methods.

\section{Conclusion and Future Work}\label{sec-concl}
Using the example of power systems transient dynamics, we presented a comparative study of different data-driven models, based on the Koopman operator theory and the graph convolutional neural network, for robust multi-timestep prediction. The Koopman-based models (namely, Robust DMD and deepDMD) require only a single time-point observation of the measurements, but need more than one type of system measurements (such as frequency and voltage magnitude). The graph convolutional neural network model (namely, STGCN), on the other hand, uses an additional information on the network topology and needs a history of observations (e.g., frequency) to simultaneously exploit the spatio-temporal and topological features of the system. Using IEEE 68-bus network as example, synthetic training and testing datasets were created (and made publicly accessible) by simulating a wide range of load change scenarios, categorized by the magnitudes of load changes, the degree of connectivity of the load change location, and the distance of the load change location to nearest generator nodes. Numerical studies were performed to evaluate and verify the performance of the models with respect to the robustness to noisy measurements, accuracy across spatio-temporal scales, sensitivity to hyper-parameters and computational time costs. This has been a preliminary investigation of two classes of complementary methods for learning (and predicting) nonlinear dynamical system behavior, using power systems as an example. Future studies will involve combining the best of the Koopman-based and graph convolution neural network models, along with distributed learning techniques, to enable robust models for large-scale dynamical networks. 



\begin{thebibliography}{10}
\providecommand{\url}[1]{#1}
\csname url@samestyle\endcsname
\providecommand{\newblock}{\relax}
\providecommand{\bibinfo}[2]{#2}
\providecommand{\BIBentrySTDinterwordspacing}{\spaceskip=0pt\relax}
\providecommand{\BIBentryALTinterwordstretchfactor}{4}
\providecommand{\BIBentryALTinterwordspacing}{\spaceskip=\fontdimen2\font plus
\BIBentryALTinterwordstretchfactor\fontdimen3\font minus
  \fontdimen4\font\relax}
\providecommand{\BIBforeignlanguage}[2]{{%
\expandafter\ifx\csname l@#1\endcsname\relax
\typeout{** WARNING: IEEEtran.bst: No hyphenation pattern has been}%
\typeout{** loaded for the language `#1'. Using the pattern for}%
\typeout{** the default language instead.}%
\else
\language=\csname l@#1\endcsname
\fi
#2}}
\providecommand{\BIBdecl}{\relax}
\BIBdecl

\bibitem{sobbouhi2021transient}
A.~R. Sobbouhi and A.~Vahedi, ``Transient stability prediction of power system;
  a review on methods, classification and considerations,'' \emph{Electric
  Power Systems Research}, vol. 190, p. 106853, 2021.

\bibitem{glavic2017reinforcement}
M.~Glavic, R.~Fonteneau, and D.~Ernst, ``Reinforcement learning for electric
  power system decision and control: Past considerations and perspectives,''
  \emph{IFAC-PapersOnLine}, vol.~50, no.~1, pp. 6918--6927, 2017.

\bibitem{huang2019adaptive}
Q.~Huang, R.~Huang, W.~Hao, J.~Tan, R.~Fan, and Z.~Huang, ``Adaptive power
  system emergency control using deep reinforcement learning,'' \emph{IEEE
  Transactions on Smart Grid}, vol.~11, no.~2, pp. 1171--1182, 2019.

\bibitem{legaard2021constructing}
C.~M. Legaard, T.~Schranz, G.~Schweiger, J.~Drgo{\v{n}}a, B.~Falay, C.~Gomes,
  A.~Iosifidis, M.~Abkar, and P.~G. Larsen, ``Constructing neural network-based
  models for simulating dynamical systems,'' \emph{arXiv preprint
  arXiv:2111.01495}, 2021.

\bibitem{wang2017new}
Y.~Wang, ``A new concept using lstm neural networks for dynamic system
  identification,'' in \emph{2017 American Control Conference (ACC)}.\hskip 1em
  plus 0.5em minus 0.4em\relax IEEE, 2017, pp. 5324--5329.

\bibitem{ogunmolu2016nonlinear}
O.~Ogunmolu, X.~Gu, S.~Jiang, and N.~Gans, ``Nonlinear systems identification
  using deep dynamic neural networks,'' \emph{arXiv preprint arXiv:1610.01439},
  2016.

\bibitem{bahbah2004new}
A.~G. Bahbah and A.~A. Girgis, ``New method for generators' angles and angular
  velocities prediction for transient stability assessment of multimachine
  power systems using recurrent artificial neural network,'' \emph{IEEE
  Transactions on Power Systems}, vol.~19, no.~2, pp. 1015--1022, 2004.

\bibitem{li2020machine}
J.~Li, M.~Yue, Y.~Zhao, and G.~Lin, ``Machine-learning-based online transient
  analysis via iterative computation of generator dynamics,'' in \emph{2020
  IEEE International Conference on Communications, Control, and Computing
  Technologies for Smart Grids (SmartGridComm)}.\hskip 1em plus 0.5em minus
  0.4em\relax IEEE, 2020, pp. 1--6.

\bibitem{liu2019transient}
L.~Liu, Y.~Li, Y.~Cao, F.~Liu, W.~Wang, and J.~Zuo, ``Transient rotor angle
  stability prediction based on deep belief network and long short-term memory
  network,'' \emph{IFAC-PapersOnLine}, vol.~52, no.~4, pp. 176--181, 2019.

\bibitem{wu2020model}
Y.~Wu, \emph{Model Parameter Calibration in Power Systems}.\hskip 1em plus
  0.5em minus 0.4em\relax MS Thesis, The University of Vermont and State
  Agricultural College, 2020.

\bibitem{zhao2020roles}
J.~Zhao, M.~Netto, Z.~Huang, S.~S. Yu, A.~Gomez-Exposito, S.~Wang, I.~Kamwa,
  S.~Akhlaghi, L.~Mili, V.~Terzija \emph{et~al.}, ``Roles of dynamic state
  estimation in power system modeling, monitoring and operation,'' \emph{IEEE
  Transactions on Power Systems}, vol.~36, no.~3, pp. 2462--2472, 2020.

\bibitem{koopman1931hamiltonian}
B.~O. Koopman, ``{Hamiltonian systems and transformation in Hilbert space},''
  \emph{Proceedings of the National Academy of Sciences}, vol.~17, no.~5, pp.
  315--318, 1931.

\bibitem{rowley2009spectral}
C.~W. Rowley, I.~Mezi{\'c}, S.~Bagheri, P.~Schlatter, and D.~S. Henningson,
  ``Spectral analysis of nonlinear flows,'' \emph{Journal of fluid mechanics},
  vol. 641, pp. 115--127, 2009.

\bibitem{tu2013dynamic}
J.~H. Tu, C.~W. Rowley, D.~M. Luchtenburg, S.~L. Brunton, and J.~N. Kutz, ``On
  dynamic mode decomposition: Theory and applications,'' \emph{Journal of
  Computational Dynamics}, vol.~1, no.~2, pp. 391--421, 2014.

\bibitem{williams2015data}
M.~O. Williams, I.~G. Kevrekidis, and C.~W. Rowley, ``{A data--driven
  approximation of the Koopman operator: Extending dynamic mode
  decomposition},'' \emph{Journal of Nonlinear Science}, vol.~25, no.~6, pp.
  1307--1346, 2015.

\bibitem{sinha2018robust}
S.~Sinha, B.~Huang, and U.~Vaidya, ``{Robust approximation of Koopman operator
  and prediction in random dynamical systems},'' in \emph{2018 Annual American
  Control Conference (ACC)}.\hskip 1em plus 0.5em minus 0.4em\relax IEEE, 2018,
  pp. 5491--5496.

\bibitem{takeishi2017learning}
N.~Takeishi, Y.~Kawahara, and T.~Yairi, ``{Learning Koopman invariant subspaces
  for dynamic mode decomposition},'' in \emph{Advances in Neural Information
  Processing Systems}, 2017, pp. 1130--1140.

\bibitem{li2017extended}
Q.~Li, F.~Dietrich, E.~M. Bollt, and I.~G. Kevrekidis, ``{Extended dynamic mode
  decomposition with dictionary learning: A data-driven adaptive spectral
  decomposition of the Koopman operator},'' \emph{Chaos: An Interdisciplinary
  Journal of Nonlinear Science}, vol.~27, no.~10, p. 103111, 2017.

\bibitem{lusch2018deep}
B.~Lusch, J.~N. Kutz, and S.~L. Brunton, ``Deep learning for universal linear
  embeddings of nonlinear dynamics,'' \emph{Nature communications}, vol.~9,
  no.~1, p. 4950, 2018.

\bibitem{yeung2019learning}
E.~Yeung, S.~Kundu, and N.~Hodas, ``{Learning deep neural network
  representations for Koopman operators of nonlinear dynamical systems},'' in
  \emph{2019 American Control Conference (ACC)}.\hskip 1em plus 0.5em minus
  0.4em\relax IEEE, 2019, pp. 4832--4839.

\bibitem{nandanoori2020data}
S.~P. Nandanoori, S.~Sinha, and E.~Yeung, ``Data-driven operator theoretic
  methods for global phase space learning,'' in \emph{2020 American Control
  Conference (ACC)}.\hskip 1em plus 0.5em minus 0.4em\relax IEEE, 2020, pp.
  4551--4557.

\bibitem{sinha2020koopman}
S.~Sinha, S.~P. Nandanoori, and E.~Yeung, ``Koopman operator methods for global
  phase space exploration of equivariant dynamical systems,''
  \emph{IFAC-PapersOnLine}, vol.~53, no.~2, pp. 1150--1155, 2020.

\bibitem{eisenhower2010decomposing}
B.~Eisenhower, T.~Maile, M.~Fischer, and I.~Mezi{\'c}, ``{Decomposing building
  system data for model validation and analysis using the Koopman operator},''
  \emph{Proceedings of SimBuild}, vol.~4, no.~1, pp. 434--441, 2010.

\bibitem{barocio2014dynamic}
E.~Barocio, B.~C. Pal, N.~F. Thornhill, and A.~R. Messina, ``A dynamic mode
  decomposition framework for global power system oscillation analysis,''
  \emph{IEEE Transactions on Power Systems}, vol.~30, no.~6, pp. 2902--2912,
  2014.

\bibitem{susuki2016applied}
Y.~Susuki, I.~Mezic, F.~Raak, and T.~Hikihara, ``Applied koopman operator
  theory for power systems technology,'' \emph{Nonlinear Theory and Its
  Applications, IEICE}, vol.~7, no.~4, pp. 430--459, 2016.

\bibitem{nandanoori2020model}
S.~P. Nandanoori, S.~Kundu, S.~Pal, K.~Agarwal, and S.~Choudhury,
  ``{Model-agnostic algorithm for real-time attack identification in power grid
  using Koopman modes},'' in \emph{2020 IEEE International Conference on
  Communications, Control, and Computing Technologies for Smart Grids
  (SmartGridComm)}.\hskip 1em plus 0.5em minus 0.4em\relax IEEE, 2020, pp.
  1--6.

\bibitem{surana2018koopman}
A.~Surana, ``Koopman operator framework for time series modeling and
  analysis,'' \emph{Journal of Nonlinear Science}, pp. 1--34, 2018.

\bibitem{sinha2020data}
S.~Sinha, S.~P. Nandanoori, and E.~Yeung, ``Data driven online learning of
  power system dynamics,'' in \emph{2020 IEEE Power \& Energy Society General
  Meeting (PESGM)}.\hskip 1em plus 0.5em minus 0.4em\relax IEEE, 2020, pp.
  1--5.

\bibitem{netto2021analytical}
M.~Netto, Y.~Susuki, V.~Krishnan, and Y.~Zhang, ``On analytical construction of
  observable functions in extended dynamic mode decomposition for nonlinear
  estimation and prediction,'' in \emph{2021 American Control Conference
  (ACC)}.\hskip 1em plus 0.5em minus 0.4em\relax IEEE, 2021, pp. 4190--4195.

\bibitem{korda2018power}
M.~Korda, Y.~Susuki, and I.~Mezi{\'c}, ``Power grid transient stabilization
  using koopman model predictive control,'' \emph{IFAC-PapersOnLine}, vol.~51,
  no.~28, pp. 297--302, 2018.

\bibitem{scarselli2008graph}
F.~Scarselli, M.~Gori, A.~C. Tsoi, M.~Hagenbuchner, and G.~Monfardini, ``The
  graph neural network model,'' \emph{IEEE transactions on neural networks},
  vol.~20, no.~1, pp. 61--80, 2008.

\bibitem{zhou2020graph}
J.~Zhou, G.~Cui, S.~Hu, Z.~Zhang, C.~Yang, Z.~Liu, L.~Wang, C.~Li, and M.~Sun,
  ``Graph neural networks: A review of methods and applications,'' \emph{AI
  Open}, vol.~1, pp. 57--81, 2020.

\bibitem{lecun2015deep}
Y.~LeCun, Y.~Bengio, and G.~Hinton, ``Deep learning,'' \emph{nature}, vol. 521,
  no. 7553, 2015.

\bibitem{yu2017spatio}
B.~Yu, H.~Yin, and Z.~Zhu, ``Spatio-temporal graph convolutional networks: A
  deep learning framework for traffic forecasting,'' \emph{International Joint
  Conference on Artificial Intelligence (IJCAI)}, 2018.

\bibitem{lin2021spatial}
W.~Lin, D.~Wu, and B.~Boulet, ``Spatial-temporal residential short-term load
  forecasting via graph neural networks,'' \emph{IEEE Transactions on Smart
  Grid}, vol.~12, no.~6, pp. 5373--5384, 2021.

\bibitem{gridstage2020}
\BIBentryALTinterwordspacing
``{GridSTAGE: SpatioTemporal Adversarial Scenario Generation framework},'' Mar.
  2020. [Online]. Available: \url{https://github.com/pnnl/GridSTAGE}
\BIBentrySTDinterwordspacing

\bibitem{nandanoori2021nominal}
S.~P. Nandanoori, ``{Nominal and adversarial synthetic PMU data for standard
  IEEE test systems},'' Pacific Northwest National Lab.(PNNL), Richland, WA
  (United States), Tech. Rep., 2021.

\bibitem{gridprediction2021}
\BIBentryALTinterwordspacing
``{Evaluation of Deep Learning Based Predictive Models Using Power Systems Use
  Case},'' Apr. 2021. [Online]. Available:
  \url{https://github.com/pnnl/grid_prediction}
\BIBentrySTDinterwordspacing

\bibitem{mezic2013analysis}
I.~Mezi{\'c}, ``{Analysis of fluid flows via spectral properties of the Koopman
  operator},'' \emph{Annual Review of Fluid Mechanics}, vol.~45, pp. 357--378,
  2013.

\bibitem{lasota2013chaos}
A.~Lasota and M.~C. Mackey, \emph{Chaos, fractals, and noise: stochastic
  aspects of dynamics}.\hskip 1em plus 0.5em minus 0.4em\relax Springer Science
  \& Business Media, 2013, vol.~97.

\bibitem{johnson2018class}
C.~A. Johnson and E.~Yeung, ``{A class of logistic functions for approximating
  state-inclusive Koopman operators},'' in \emph{2018 Annual American Control
  Conference (ACC)}.\hskip 1em plus 0.5em minus 0.4em\relax IEEE, 2018, pp.
  4803--4810.

\bibitem{shuman2013emerging}
D.~I. Shuman, S.~K. Narang, P.~Frossard, A.~Ortega, and P.~Vandergheynst, ``The
  emerging field of signal processing on graphs: Extending high-dimensional
  data analysis to networks and other irregular domains,'' \emph{IEEE signal
  processing magazine}, vol.~30, no.~3, 2013.

\bibitem{defferrard2016convolutional}
M.~Defferrard, X.~Bresson, and P.~Vandergheynst, ``Convolutional neural
  networks on graphs with fast localized spectral filtering,'' in
  \emph{NeurIPS}, 2016, pp. 3844--3852.

\bibitem{kipf2016semi}
T.~N. Kipf and M.~Welling, ``Semi-supervised classification with graph
  convolutional networks,'' \emph{arXiv preprint arXiv:1609.02907}, 2016.

\bibitem{oord2016wavenet}
A.~v.~d. Oord, S.~Dieleman, H.~Zen, K.~Simonyan, O.~Vinyals, A.~Graves,
  N.~Kalchbrenner, A.~Senior, and K.~Kavukcuoglu, ``Wavenet: A generative model
  for raw audio,'' \emph{arXiv preprint arXiv:1609.03499}, 2016.

\bibitem{dauphin2017language}
Y.~N. Dauphin, A.~Fan, M.~Auli, and D.~Grangier, ``Language modeling with gated
  convolutional networks,'' in \emph{International conference on machine
  learning}.\hskip 1em plus 0.5em minus 0.4em\relax PMLR, 2017, pp. 933--941.

\bibitem{chow1992toolbox}
J.~H. Chow and K.~W. Cheung, ``A toolbox for power system dynamics and control
  engineering education and research,'' \emph{IEEE Trans. Power Systems},
  vol.~7, no.~4, pp. 1559--1564, 1992.

\bibitem{bhandari2019real}
A.~Bhandari, H.~Yin, Y.~Liu, W.~Yao, and L.~Zhan, ``Real-time signal-to-noise
  ratio estimation by universal grid analyzer,'' in \emph{2019 International
  Conference on Smart Grid Synchronized Measurements and Analytics
  (SGSMA)}.\hskip 1em plus 0.5em minus 0.4em\relax IEEE, 2019, pp. 1--6.

\end{thebibliography}
\end{document}